\documentclass[sigconf]{acmart}
\pdfoutput=1
\setcopyright{none}
\renewcommand\footnotetextcopyrightpermission[1]{}
\usepackage{subcaption}
\usepackage{booktabs} 
\usepackage{graphicx}
\usepackage{enumerate}
\usepackage{enumitem}
\usepackage[british]{babel}
\usepackage{hhline}
\usepackage{multirow}
\usepackage{hyperref}
\usepackage{caption}
\usepackage[ruled, boxed,vlined,linesnumbered]{algorithm2e}
\usepackage{xcolor}
\usepackage{xspace}
\usepackage{algorithmic}
\usepackage{dblfloatfix}
\usepackage{balance}
\usepackage[export]{adjustbox}

\newcommand{\note}[1]{\textcolor{red}{\em #1}}
\newcommand{\ignore}[1]{}
\newcommand\sigmodpagestyle{plain}

\AtBeginDocument{%
  \providecommand\BibTeX{{%
    \normalfont B\kern-0.5em{\scshape i\kern-0.25em b}\kern-0.8em\TeX}}}

\begin{document}



\title{Serverless Data Science - Are We There Yet? A Case Study of Model Serving}

\author{Yuncheng Wu}
\affiliation{%
  \institution{National University of Singapore}
  \country{Singapore}
}
\email{wuyc@comp.nus.edu.sg}
\author{Tien Tuan Anh Dinh}
\affiliation{%
  \institution{Singapore University of Technology and Design}
  \country{Singapore}
}
\email{dinhtta@sutd.edu.sg}
\author{Guoyu Hu}
\affiliation{%
  \institution{National University of Singapore}
  \country{Singapore}
}
\email{guoyu.hu@u.nus.edu}
\author{Meihui Zhang}
\affiliation{%
  \institution{Beijing Institute of Technology}
  \country{P. R. China}
}
\email{meihui\_zhang@bit.edu.cn}
\author{Yeow Meng Chee}
\affiliation{%
  \institution{National University of Singapore}
  \country{Singapore}
}
\email{ymchee@nus.edu.sg}
\author{Beng Chin Ooi}
\affiliation{%
  \institution{National University of Singapore}
  \country{Singapore}
}
\email{ooibc@comp.nus.edu.sg}

\renewcommand{\shortauthors}{Anonymous author(s).}



\begin{abstract}
Machine learning (ML) is an important part of modern data science applications. Data scientists today have to manage the end-to-end ML life cycle that includes both model training and model serving, the latter of which is essential, as it makes their works available to end-users.
Systems of model serving require high performance, low cost, and ease of management. Cloud providers are already offering model serving choices, including managed services and self-rented servers. Recently, serverless computing, whose advantages include high elasticity and a fine-grained cost model, brings another option for model serving.

Our goal in this paper is to examine the viability of serverless as a mainstream model serving platform. To this end, we first conduct a comprehensive evaluation of the performance and cost of serverless against other model serving systems on Amazon Web Service and Google Cloud Platform. We find that serverless outperforms many cloud-based alternatives. Further, there are settings under which it even achieves better performance than GPU-based systems.
%
Next, we present the design space of serverless model serving, which comprises multiple dimensions, including cloud platforms, serving runtimes, and other function-specific parameters. For each dimension, we analyze the impact of different choices and provide suggestions for data scientists to better utilize serverless model serving. Finally, we discuss challenges and opportunities in building a more practical serverless model serving system. 

\end{abstract}


\maketitle

\pagestyle{\sigmodpagestyle}

\section{Introduction}
\label{sec:introduction}

Machine learning (ML) has transformed data science~\cite{facebookml,jiang18,velikovich18}.
Figure~\ref{fig:data-science-pipeline} illustrates the changes. Until recently, a typical data science pipeline consists of data collection, cleaning and integration, data analytics, and visualization. Now, the pipeline includes the full ML life cycle: feature engineering, model training, and model serving, turning data scientists into end-to-end data engineers~\cite{mlcask2021}. 
%
%
Consider a nutrition analysis application,  FoodLG\footnote{http://www.foodlg.com/}, 
as an example. A data scientist first collects a set of labeled food images from nutritionists,
transforms the data, 
and builds a deep learning model~\cite{OoiTWWCCGLTWXZZ15, 0059Z0JOT16} that classifies food images and analyzes the nutrition according to a food knowledge base (e.g., CalorieKing~\cite{calorieking}).
%
After that, the data scientist deploys the model and makes it available to users' mobile apps. Then a user sends food images as requests to the deployed model and receives the predicted food nutrition (e.g., calorie, protein). The application can record the user's daily intake, analyze her eating habit, and provide suggestions for healthy dietary intake. 
New data is collected and fed back to the pipeline to improve model accuracy.  
In real-world applications, model serving plays a crucial role in bringing the works of data scientists to the end-users. 

\ignore{
\begin{figure}[h]
\centering
\begin{subfigure}[b]{0.2\textwidth}
    \centering
    \includegraphics[width=0.7\columnwidth]{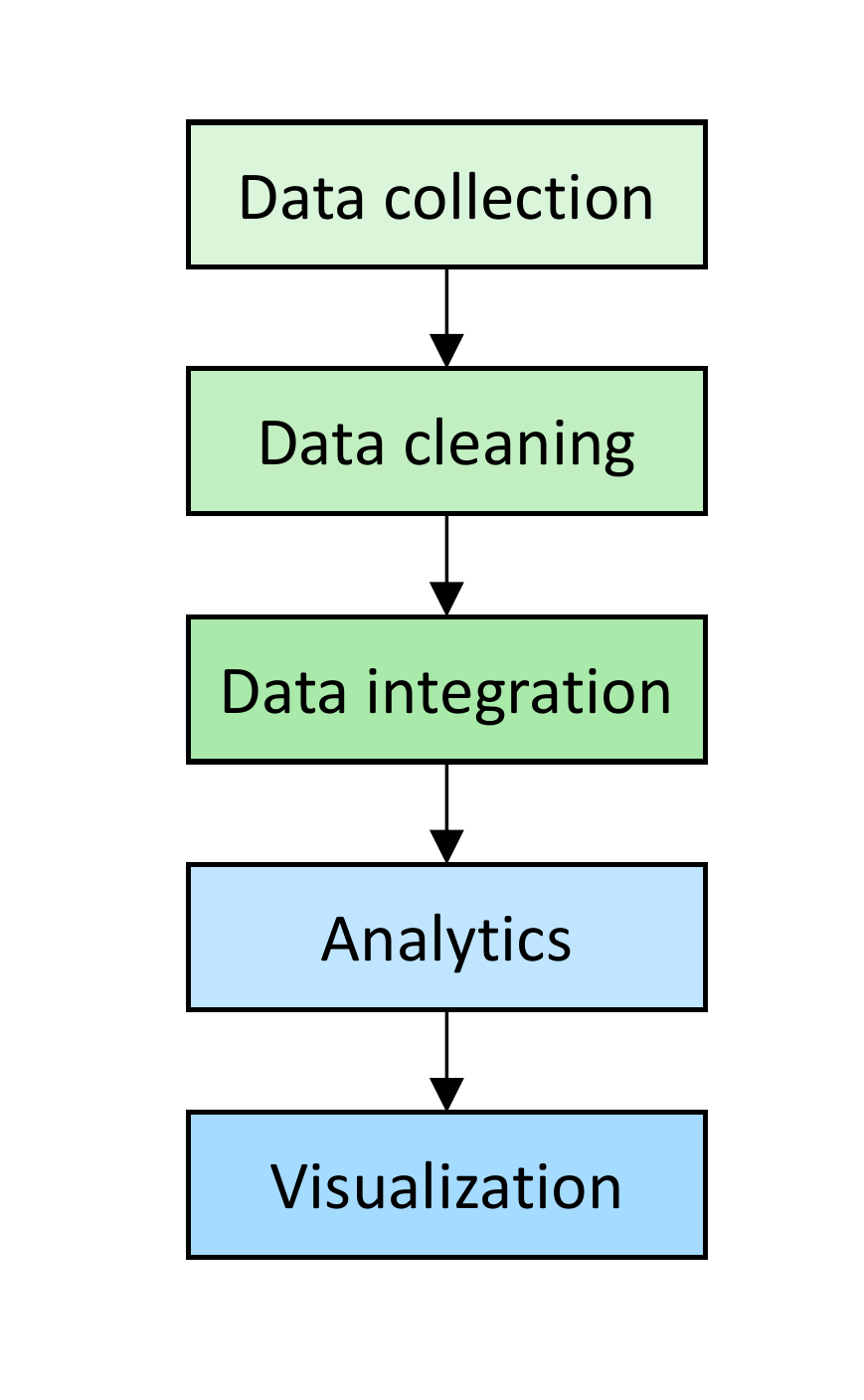}
    \vspace{-1mm}
    \caption{before machine learning}
    \label{subfig:pipeline-without-ml}
\end{subfigure}
~
\begin{subfigure}[b]{0.2\textwidth}
    \centering
    \includegraphics[width=0.7\columnwidth]{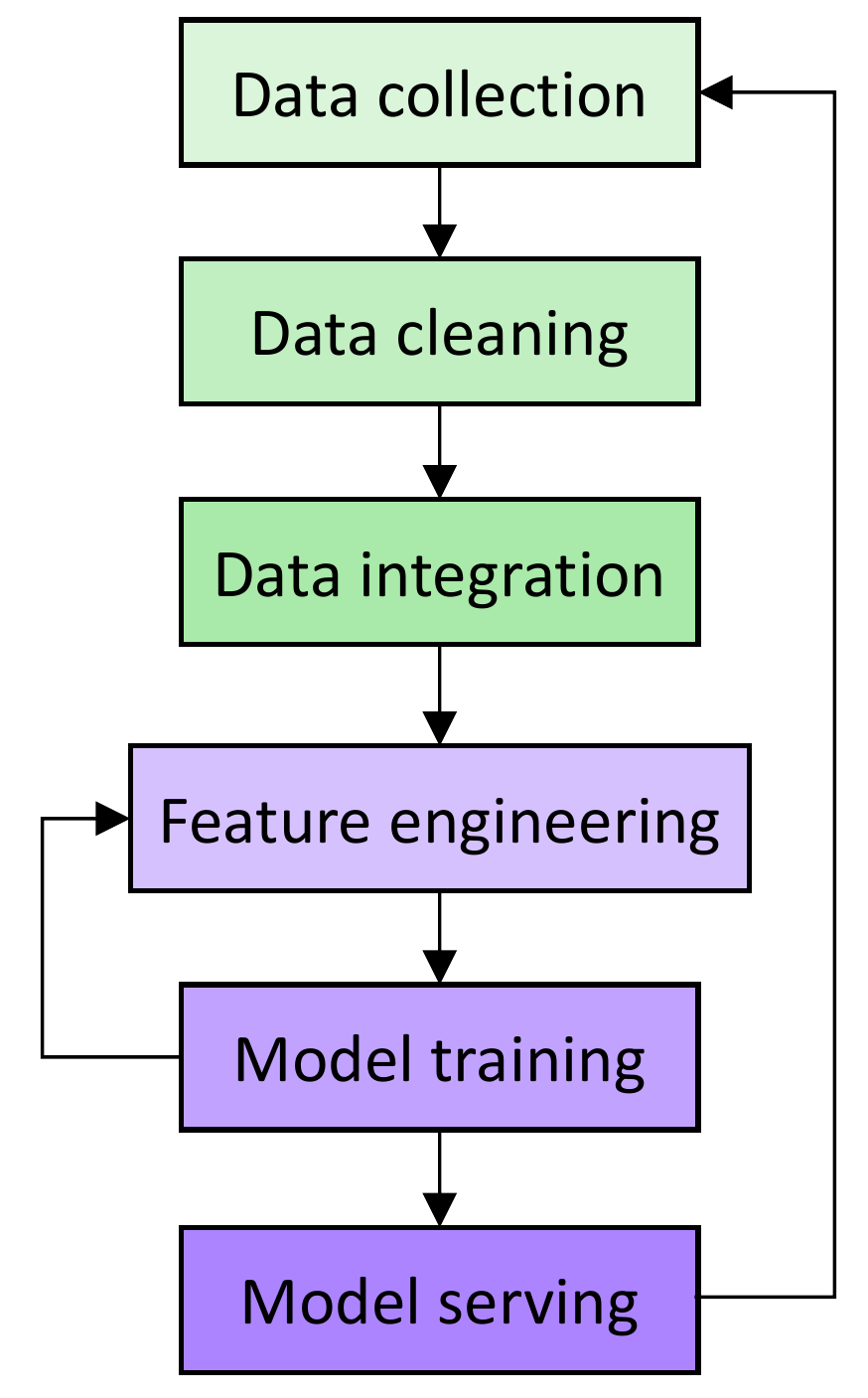}
    \vspace{-1mm}
    \caption{with machine learning}
    \label{subfig:pipeline-with-ml}
\end{subfigure}
\vspace{-3mm}
\caption{A typical data science pipeline.}
\label{fig:data-science-pipeline}
\end{figure}
\vspace{-3mm}
}

\vspace{-3mm}
\begin{figure}[h]
\centering
\begin{subfigure}[b]{0.99\columnwidth}
    \centering
    \includegraphics[width=0.99\columnwidth]{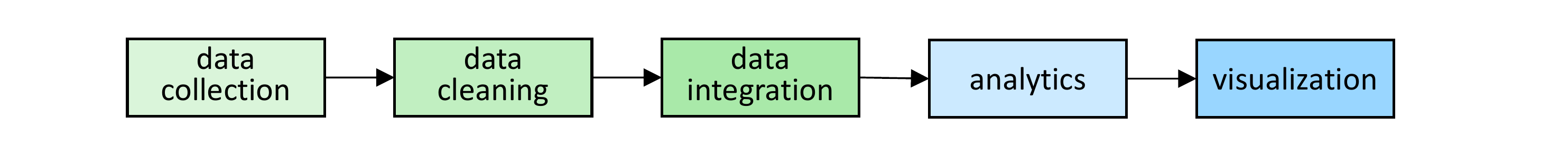}
    \vspace{-4mm}
    \caption{Before machine learning}
    \label{subfig:pipeline-without-ml}
\end{subfigure}

\begin{subfigure}[b]{0.99\columnwidth}
    \centering
    \includegraphics[width=0.99\columnwidth]{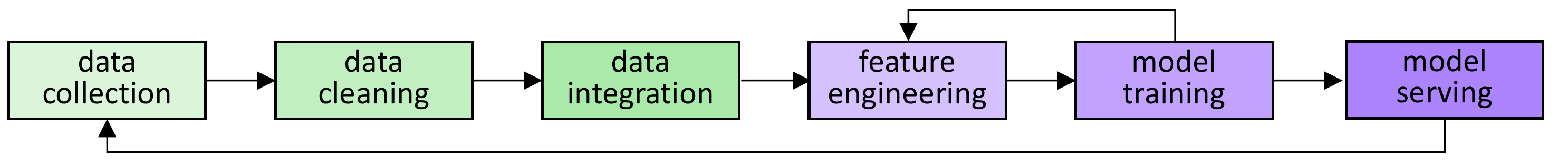}
    \vspace{-2mm}
    \caption{With machine learning}
    \label{subfig:pipeline-with-ml}
\end{subfigure}
\vspace{-4mm}
\caption{A typical data science pipeline.}
\label{fig:data-science-pipeline}
\end{figure}
\vspace{-4mm}

%

Model serving systems are interactive, that is, they handle inference requests from users in near real-time. As a result, their design goals differ from those of model training systems that focus on maximizing throughput. 
The first goal is {\em high performance}, which means that the system can process requests fast, even under bursty workloads. 
%
The second goal is {\em low cost}, which means that the system is cost-effective when handling a large number of
requests. 
The third goal is {\em ease of management}, which means that the system allows data scientists to quickly deploy ML models without worrying about low-level details such as resource management.
Existing cloud providers are offering several model serving choices, including managed ML services and self-rented servers~\cite{aws-sagemaker,gcp-ai-platform,azureml}. 
These options, however, do not meet all three goals.

%
\ignore{
\begin{figure}[t]
    \centering
    \includegraphics[width=0.4\textwidth]{figures/datascience.pdf}
    \caption{A typical data science pipeline. Left: before machine learning. Right: with 
    machine learning.}
    \label{fig:data-science-pipeline}
\end{figure}
}

\begin{figure}[t]
    \centering
    \includegraphics[width=0.4\textwidth]{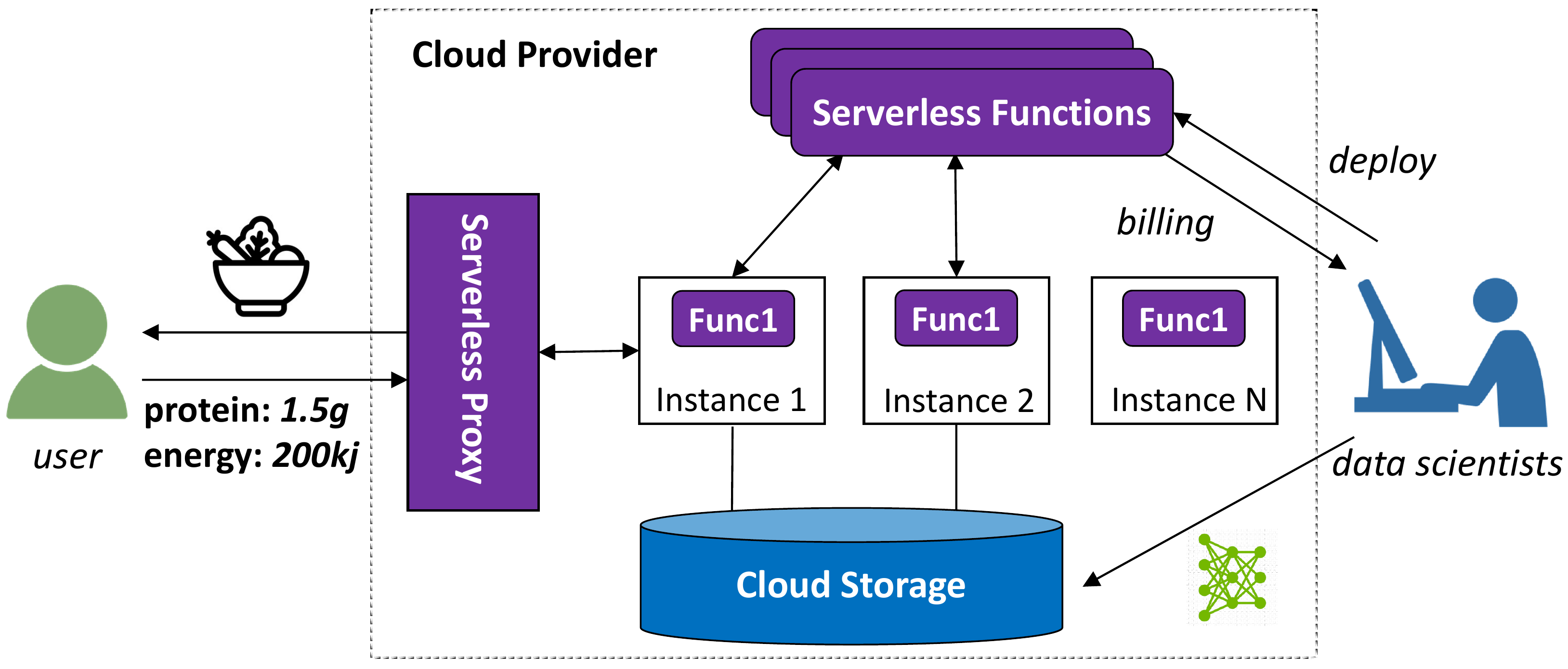}
    \vspace{-3mm}
    \caption{Serverless model serving example.}
    \vspace{-6mm}
    \label{fig:serverless-example}
\end{figure}

  
Serverless computing~\cite{hellerstein19, berkeley-serverless}, an emerging cloud computing paradigm,
brings another option for model serving, and it has the potentials to achieve the three goals above. 
Figure~\ref{fig:serverless-example} shows how serverless can support model serving for the aforementioned application. 
The data scientist first uploads a trained model to the cloud storage, then writes and deploys a function for executing model inference. 
The cloud provider creates an invokable endpoint (e.g., a URL) for the function, to which users can send their
requests and receive inference results. 
The serverless platform takes care of provisioning computation instances to execute the function, and of scaling the
number of instances according to the request workload. 
The data scientist is charged by the actual amount of consumed resources, instead of the time of reserved resources. 


Serverless can achieve the first goal of model serving, i.e., high performance, because it adaptively scales to multiple instances very fast so that it can handle the requests timely. It can meet the second goal, i.e., low cost, because billing is only based on the actual resource consumption. This means that the scientist does not have to provision for peak load, which can be expensive, especially when using GPUs. The last goal, ease of management, is also met since provisioning and scaling of resources are handled automatically by the cloud platform.
Despite these potentials, there are challenges in applying serverless to model serving. In particular, serverless has several limitations, including small memory size, limited running time, and lack of persistent states~\cite{hellerstein19, 0002MA20, jonas2017occupy, klimovic2018pocket, klimovic2018understanding, SreekantiWLSGHT20, Shafiei2019}.

In this paper, we ask the following question: {\it can serverless computing be a mainstream model serving platform for data science applications? } 
%
%
To answer this, we conduct an extensive comparison between serverless and other cloud-based model serving alternatives with respect to performance and cost.
We consider eight systems spanning two major cloud providers: Amazon Web Service~\cite{aws} (AWS) and Google Cloud Platform~\cite{gcp} (GCP).
In particular, we evaluate Lambda, Cloud Functions, SageMaker, AI Platform, and self-rented CPU and GPU systems from AWS and GCP. We use three deep learning models for the evaluation: MobileNet and VGG, two image classification models, and ALBERT, a natural language processing model. We evaluate these models on the eight systems under three different workloads. We compare the systems in three metrics: response latency, request success ratio, and cost.  
%
%



Our results contain two surprising findings. First, while earlier works claimed that serverless is not
suitable for model serving~\cite{hellerstein19}, we find that, in most cases, serverless outperforms managed ML services such as SageMaker and AI Platform in both cost and performance. Meanwhile, it has better performance than self-rented CPU systems but with a higher cost; nevertheless, serverless can be much faster if the cost is relatively comparable on AWS.
Second, while other works suggested that serverless should be used as a complementary platform to a GPU-based system for handling excessive load~\cite{mark}, we show that there are settings under which serverless could be a better choice than self-rented GPU systems. 
{
In particular, on AWS, serving MobileNet with workload-200 (see Figure~\ref{fig:workloads}, which consists of 86000 requests within 15 minutes) with serverless results in an average latency of $0.097s$ and a cost of $\$0.186$. In contrast, doing the same using a GPU server results in $7.52s$ and $\$0.187$, respectively. With comparable cost, serverless achieves 77.5$\times$ latency improvement.}
Importantly, we find that serverless is less sensitive to changes in workloads or models, which means it can provide consistent performance even under bursty workloads or given large models.
Our evaluation results provide an important insight, that is, serverless is a viable and promising option for model serving.

Next, we describe the design space of serverless model serving, which comprises multiple dimensions, including serverless platforms, serving runtimes, and other function-specific parameters. We further investigate the impact of each dimension on serverless' performance and cost, and present suggestions for data scientists to better utilize serverless model serving. 
{
In particular, we observe that there is a gap in the cold start time between AWS and GCP serverless functions. For instance, using the same serving environment, namely TensorFlow1.15~\cite{tensorflow}, AWS takes about 9.49$s$ on average to start an instance while GCP takes about 14.19$s$ when serving ALBERT with workload-120 (see Figure~\ref{fig:workloads}).}
Another interesting finding is that a lightweight serving runtime could reduce both latency and cost in most cases. 
{
For instance, if serving with OnnxRuntime1.4~\cite{onnxruntime} instead of TensorFlow1.15, we can obtain up to 3.61$\times$ latency speedup and 4.55$\times$ cost reduction. 
}

In summary, we make the following contributions in this paper.
\begin{itemize}[topsep=0pt,itemsep=0pt,parsep=0pt,partopsep=0pt,leftmargin=15pt]
\item 
We conduct an extensive cost and performance comparison of serverless against other cloud-based systems for model serving. 
Our analysis covers eight different systems on two major public cloud providers, with three deep learning models and three different workloads.
\item 
We present interesting findings indicating that serverless is a viable and promising option for model serving. 
We show that it outperforms managed ML services and CPU-based systems in most cases, and there are settings under which it can even achieve better performance than GPU-based systems.  

\item We further explore the impact of its design space on serverless model serving and present recommendations for data scientists on how to better use serverless for model serving. In addition, we discuss research challenges and opportunities in building a more practical serverless model serving system. 

\end{itemize}
%
%

\ignore{
\begin{enumerate}
\item Start-up challenge: start-up latency for a function consists of the time for booting the sandbox, and for
setting up the runtime. The later includes loading necessary libraries for running the function, which can be
high for machine learning workloads. Thus, the challenge is to minimize start-up time.
\item Security challenge: to communicate with a SGX-based server, user must establish a secure channel after
running an attestation protocol. Although this protocol has high overhead, user only runs it once for each
server. Serverless functions, however, are not bound to a fixed set of servers, meaning that the user may have
to run attestation for prior to every single request. Therefore, the challenge is to minimize attestation
overhead.
\item \note{Any other challenges?}
\end{enumerate}
}


\section{Background and Related Work}\label{sec:background}


\subsection{Machine Learning Model Serving}
\label{subsec:ml-serving}
Machine learning has shown great success in data science applications~\cite{facebookml,jiang18,velikovich18}. How to efficiently deploy the trained ML models to serve end-users with low latency becomes increasingly important.
Recently, there are a number of model serving systems~\cite{clipper, rafiki, swayam} that focus on improving cost-effectiveness or model accuracy while meeting service level objective (SLO) on latency, by automatically selecting different configurations.
%
%
However, these systems are mainly based on server machines and do not work well for ML model serving due to the mismatch of target workloads~\cite{mark}.

\subsection{Serverless Computing}
\label{subsec:serverless-computing}

Serverless computing~\cite{berkeley-serverless, DukicBSA20, Ustiugov0KBG21, FuerstS21} is a recent rise of cloud computing execution paradigm such that the cloud provider runs the server and dynamically manages the allocation of resources. 
%
%
%
%
%
%
It can simplify the deployment process of function code to the production stage. 
Meanwhile, it can automatically increase and release resources to adapt to the number of user invocations, making it elastic to handle various workloads.
Pricing is based on the actual amount of resources consumed by the deployed function.
Due to its high elasticity and fine-grained cost model, serverless computing has been adopted in many data science applications, such as database analytics~\cite{jonas2017occupy, abs-1810-09679, locus, 0002MA20, PerronFDM20, Wawrzoniak0AB21}, and model training~\cite{FengKSH18, cirrus, WangNL19, JiangGLWAKS0021}.

\subsection{Serverless Model Serving}\label{subsec:background-serverless-serving}

In particular, serverless computing can be seamlessly utilized for model serving due to its stateless computations~\cite{GujaratiKAHKVM20}. As mentioned in Section~\ref{sec:introduction}, data scientists (aka. function developers) can deploy a pre-trained model on the cloud via a function. 
%
%
For a deployed function, once received an {\it event} from the serverless proxy, the cloud provider will create a new instance or forward {\it event} to a warmed instance for execution.
{
If an instance is newly created, it imports the serving dependencies, downloads the model (which was uploaded by the data scientists) from cloud storage, and loads the model into the serving runtime. Otherwise, the loaded model already exists.}
As a result, the instance parses the input sample from {\it event}, executes the inference, and returns the prediction.

\ignore{
\IncMargin{0.5em}
\begin{algorithm}[t]
\SetKwFunction{algo}{algo}
\SetKwFunction{proc}{{ServerlessHandler}}
\SetKwProg{myalg}{Algorithm}{}{}
\SetKwProg{myproc}{Procedure}{}{}
\DontPrintSemicolon
\small
\KwIn{{\it bucket}: the URL of uploaded model in cloud storage
}
\KwOut{{\it prediction}: the model inference result}
{
{\it model} = {\it None}  \\
\myproc{\proc{event}}{
  \textbf{global} {\it model} \\
  \If{model is None}{
    {\it model} $\leftarrow$ \texttt{DownloadModel}({\it bucket}) \\
  }
  {\it prediction} $\leftarrow$ \texttt{ModelInference}({\it model}, {\it event}) \\
  \KwRet{prediction} \;
}
}
\caption{Serverless model serving}
\label{alg:serverless-serving}
\end{algorithm}
}

The most related work to ours is MArK~\cite{mark}, which mainly evaluates the cost of several model serving systems on the cloud and proposes a serving system that combines self-rented servers and serverless to reduce the cost. 
BATCH~\cite{BATCH-serve} also considers serverless model serving, which designs a mechanism that batches requests and dynamically chooses the resource configuration of invoked serverless function, to satisfy latency constraints and minimize cost. 
However, neither of them has a comprehensive investigation of the performance of existing serverless systems for model serving.
%
%

{
Several other works~\cite{MaissenFKS20, YuLDXZLYQ020, copik2021sebs} explore the characteristics of serverless platforms based on various applications (e.g., image resizing). In essence, these applications follow the same stateless execution paradigm as model serving, which reads the data, performs pre-defined calculations, and returns the results. However, they do not cover two specific characteristics of serverless model serving: it requires a much longer provisioning time to prepare for the serving environment~\cite{swayam}, and model serving workloads are often unpredictable and highly bursty~\cite{mark}. 
%
}

\subsection{Model Serving Systems on the Cloud}\label{subsec:systems}


\noindent
\textbf{Serverless model serving systems.} 
We consider two serverless platforms: {\it AWS Lambda}~\cite{aws-lambda} and {\it Google Cloud Functions} (CF)~\cite{gcp-cf}.
On Lambda, model owners first package the serving environment (e.g., TensorFlow1.15) into a zip file or a container image and upload it to the cloud storage. Then, they create a function by specifying the configuration (e.g., memory size) for each instance and deploy the function based on the uploaded serving environment. We adopt the container image method in our experiments, where the image size is limited to 10GB. 
%
%
%
%
%
On CF, we can only specify the required environment in the \textit{Requirement.txt}, and the platform will build the package automatically. 
Model owners are charged by the amount of consumed resources, which mainly depends on the selected memory size and the number of instances created at runtime.
%

\noindent
\textbf{Managed machine learning (ManagedML) services.} 
Most cloud providers offer fully managed services for scientists to build, train, and deploy ML models easily.
These ManagedML services are naturally applicable for model serving, with the ability to autoscale when the workload is too heavy to be processed by existing instances.
%
%
In the experiments, we evaluate two ManagedML services: {\it AWS SageMaker}~\cite{aws-sagemaker} and {\it Google AI Platform}~\cite{gcp-ai-platform}.
On both services, model owners can create an endpoint by uploading a pre-trained model to the cloud storage and specifying a serving runtime to execute that model. 
%
{
Available frameworks on SageMaker include TensorFlow, PyTorch, MXNet, SciKit Learn, and so on. While on AI Platform, it only supports TensorFlow, XGBoost, and SciKit Learn. 
}
%
The cost is computed based on the total execution time of active instances.
%
%
%

\noindent
\textbf{Self-rented servers for model serving.} 
With self-rented servers, model owners can run virtual machines (VMs) with various configurations (e.g., vCPUs, memory, network, and storage) and deploy a model serving service by themselves on the rented instances. 
Cloud providers also provide flexible pricing options to the customers. 
%
In this paper, we deploy model serving services on both CPU servers and GPU servers on {\it AWS EC2}~\cite{aws-ec2} and {\it Google Compute Engine}~\cite{gcp-ce}.

{
Unless stated otherwise, we use AWS-Serverless, GCP-Serverless, AWS-ManagedML, and GCP-ManagedML to denote Lambda, Cloud Functions, SageMaker, and AI Platform, for better presentation.
}

\section{Evaluation Methodology}\label{sec:evaluation-method}

\begin{figure}[t]
    \centering
    \includegraphics[width=0.45\textwidth]{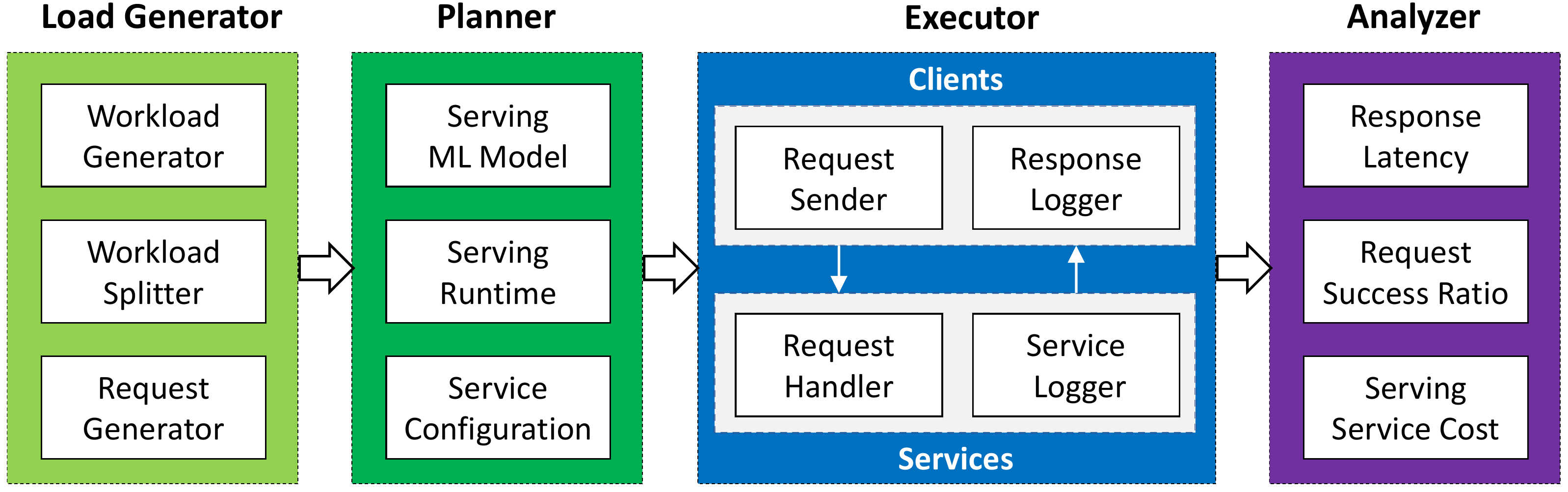}
    \vspace{-3mm}
\caption{Evaluation framework.}
\vspace{-3mm}
\label{fig:framework}
\end{figure}

We design a benchmarking framework for understanding and comparing the performance of different model serving systems. It consists of four main components as shown in Figure~\ref{fig:framework}, namely a load generator, a planner, an executor, and an analyzer. The framework can be deployed over multiple nodes. It comes with default workloads and configuration, and it can be easily extended to support new models and new platforms. 

\ignore{
\begin{table}[b]
\centering
\scriptsize
\caption{Evaluation dimensions on the cloud platforms}
\vspace{-3mm}
\centering
\begin{tabular}{|| c | c | c | c | c || }
\hline
\multicolumn{2}{||c|}{{\textbf{Services}}} & {\textbf{Instance Types}} & {\textbf{Models}} & {\textbf{Workloads}} \\
\hline
\hline
\multirow{4}{*}{AWS} & Lambda & 2GB memory & ALL & 40,120,200 \\
& SageMaker &  ml.m4.2xlarge (8vCPUs and 32GB) & ALL & 40,120 \\
& CPU Server & t2.2xlarge (8vCPUs and 32GB) & ALL & 40,120,200 \\
& GPU Server & g4dn.2xlarge (8vCPUs and 32GB) & ALL & 40,120,200 \\
\hline
\multirow{4}{*}{GCP} & Cloud Function & 2GB memory & ALL & 40,120,200 \\
& AI Platform & n1-standard-8 (8vCPUs and 30GB) & ALL & 40,120 \\
& CPU Server & n1-standard-8 (8vCPUs and 30GB) & ALL & 40,120,200 \\
& GPU Server & n1-standard-8 with 1 Tesla® T4 & ALL & 40,120,200 \\
\hline
\end{tabular}
\label{table:google-amazon-exps}
\end{table}
}

\noindent
\textbf{Load generator.}
At first, we generate workloads and requests for the clients. 
{
We note that there are no publicly available workloads for model serving~\cite{mark}. 
%
Therefore, we use the Markov-Modulated Poisson Process (MMPP) model~\cite{FischerM93, RajabiW12}, also adopted in~\cite{mark, BATCH-serve}, to generate synthetic workloads with different arrival rates and numbers of requests. The generated workloads are highly unpredictable as the occurrence and duration of demand surges are random.
}

We implement a workload splitter to evenly divide the workloads such that we can employ multiple clients to send requests, and the aggregated request rate matches the original workloads. 
Besides, we implement a request generator that creates a pool of requests, from which a client randomly selects one request to send, ensuring that model serving systems do not cache the prediction results.

\begin{figure}[t]
    \centering
    \includegraphics[width=0.28\textwidth]{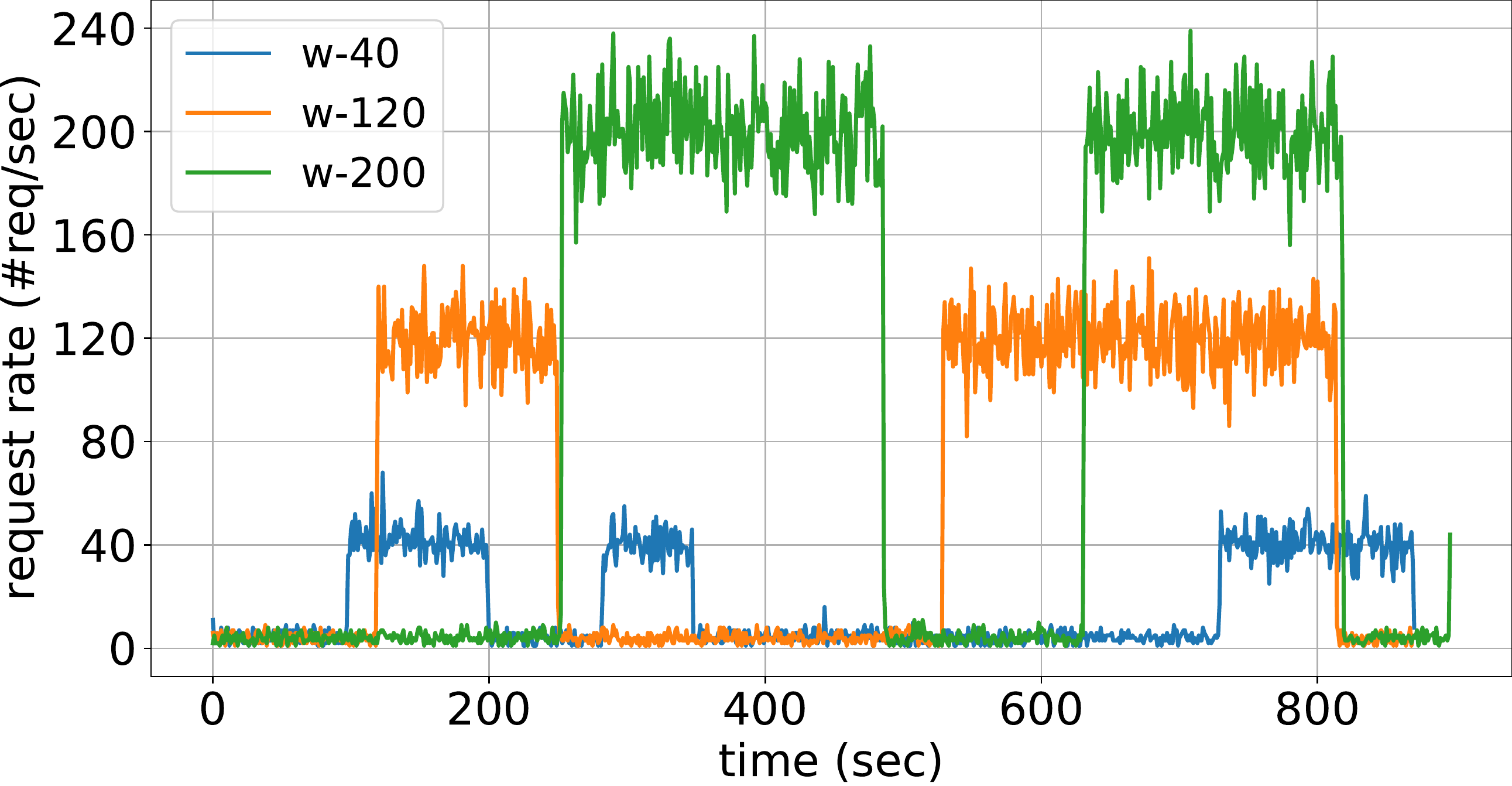}
    \vspace{-3mm}
\caption{Generated MMPP workloads.}
\vspace{-3mm}
\label{fig:workloads}
\end{figure}

In this paper, we generate three workloads to simulate low, medium, and high request rates, as shown in Figure~\ref{fig:workloads}. The numbers 40, 120, and 200 in the workloads represent the higher arrival rate of the two Poisson processes we used in MMPP. 
{
Specifically, we refer to the synthetic workloads in~\cite{mark, BATCH-serve} to select the higher arrival rates. 
}
Meanwhile, the 40, 120, and 200 workloads consist of 15000, 51600, and 86000 requests, respectively (duration is about 15 minutes).
Besides, the workloads are split for 8 clients, and the request pool size is set to 200.

\noindent
\textbf{Planner.} Before serving clients' requests, we need to deploy the serving services on the cloud. 
Specifically, the deployment is based on three dimensions: model, runtime, and configuration, which exactly define the environment on the instances to serve the requests.  

In this paper, we consider two serving runtimes: TensorFlow (TF) 1.15 and OnnxRuntime (ORT) 1.4. 
{
The first is used for a fair comparison between serverless and other serving systems, as it is one of the most popular deep learning frameworks and is well-supported by ManagedML service and self-rented servers on both clouds. In particular, GCP-ManagedML only supports TF for deep learning. The second runtime (with smaller runtime size and optimized executions) is used for exploring the performance improvement on serverless platforms, as will be discussed in Section~\ref{subsec:runtime}.
}

Meanwhile, we use three deep learning models representing two popular data science applications: MobileNet~\cite{HowardZCKWWAA17} and VGG~\cite{SimonyanZ14a} for image classification, and ALBERT~\cite{LanCGGSS20} for natural language processing. The model sizes are 16MB, 51.5MB, and 548MB, respectively.
%
{A noteworthy aspect is that there is a temporary directory storage limitation (i.e., 512MB) on AWS-Serverless~\cite{aws-lambda-quotas}, which means we cannot download the VGG model from cloud storage as the other two models. Therefore, we directly pack the VGG model into another directory inside the uploaded image for serving this model. 
%
By default, we use 2GB memory for both serverless platforms. We apply ml.m4.2xlarge (with 8vCPUs and 32GB memory) on AWS-ManagedML and n1-standard-8 (with 8vCPUs and 30GB memory) on GCP-ManagedML. Besides, we use similar configurations for self-rented CPU servers, while for GPU servers, we use g4dn.2xlarge and n1-standard-8 with 1 Tesla T4 on AWS and GCP, respectively.}

\noindent
\textbf{Executor.} 
After a serving service is deployed, the clients can send requests to the service according to the workloads. Specifically, each client randomly picks one request from the pool, sends it through designated APIs, and receives the response. We parse the response and record it into a log file, including response status and latency. For serverless serving and self-rented servers, we use standard HTTP API to send the requests. While for ML services, requests need to be sent through the API provided by the cloud platform, e.g., Amazon Boto3 and Google API Client.

When receiving a client's request, the model serving systems allocate necessary resources to process the request, including pre-processing the request (e.g., resizing the image, downloading the required model), computing the model prediction, and retrieving the label, before responding to the client. The cloud systems also record the execution logs for further analysis.

\noindent
\textbf{Analyzer.} After finishing the model serving for the workloads, we collect the results from both clients and cloud serving services, and analyze the systems using three main metrics. 
\begin{itemize}[topsep=0pt,itemsep=0pt,parsep=0pt,partopsep=0pt,leftmargin=15pt]
    \item {Response latency.} We analyze the end-to-end response latency for each request on the clients, report the average latency of the successful requests, and measure the latency trends with respect to timestamps in the workloads. 
    \item {Request success ratio (SR).} When the request rate exceeds the limit that a system can handle, some requests are dropped, or errors are returned. We analyze the ratio of successful requests over all the requests. The higher the ratio, the better. 
    \item {Cost.} We analyze the charges for each experiment. For systems charged hourly (e.g., self-rented servers), we estimate the cost based on the actual execution time. 
\end{itemize}
\section{Model Serving Systems Comparison}
\label{sec:google-amazon-exp}

{
\begin{figure}[t]
\centering
\begin{subfigure}[b]{0.95\columnwidth}
    \hspace{0.3mm}
    \includegraphics[width=0.98\columnwidth]{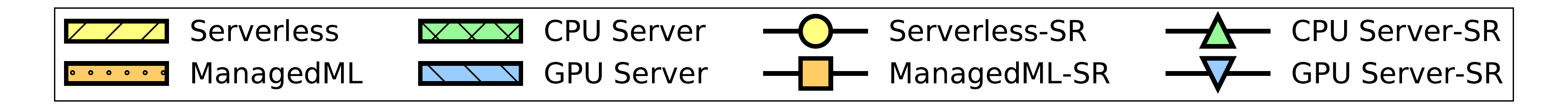}
\end{subfigure}

\centering
\begin{subfigure}[b]{0.48\columnwidth}
    \centering
    \includegraphics[width=0.98\columnwidth]{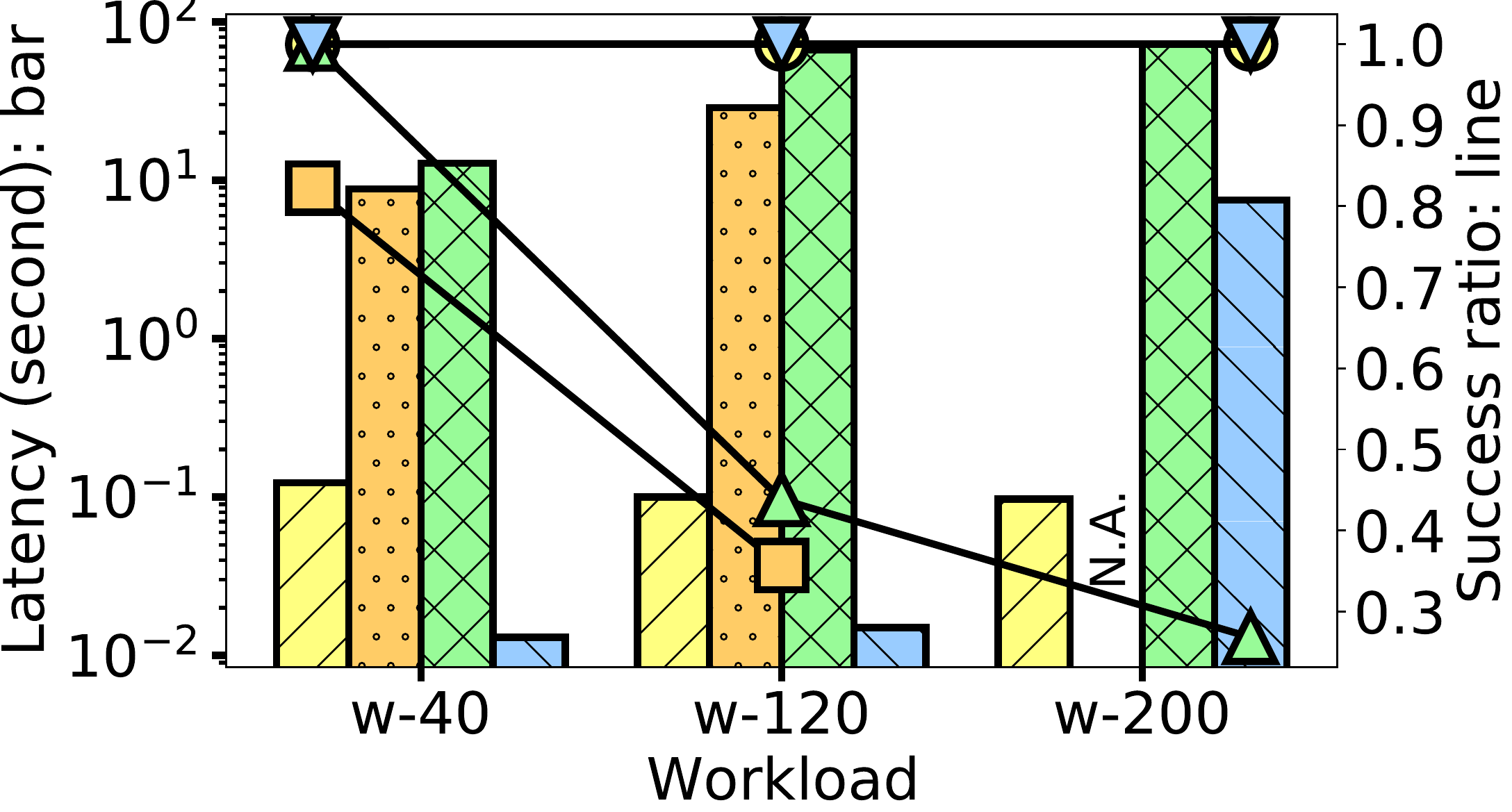}
    \vspace{-2mm}
    \caption{MobileNet (AWS)}
    \vspace{-1mm}
    \label{subfig:aws-mobilenet}
\end{subfigure}
~
\begin{subfigure}[b]{0.48\columnwidth}
    \centering
    \includegraphics[width=0.98\columnwidth]{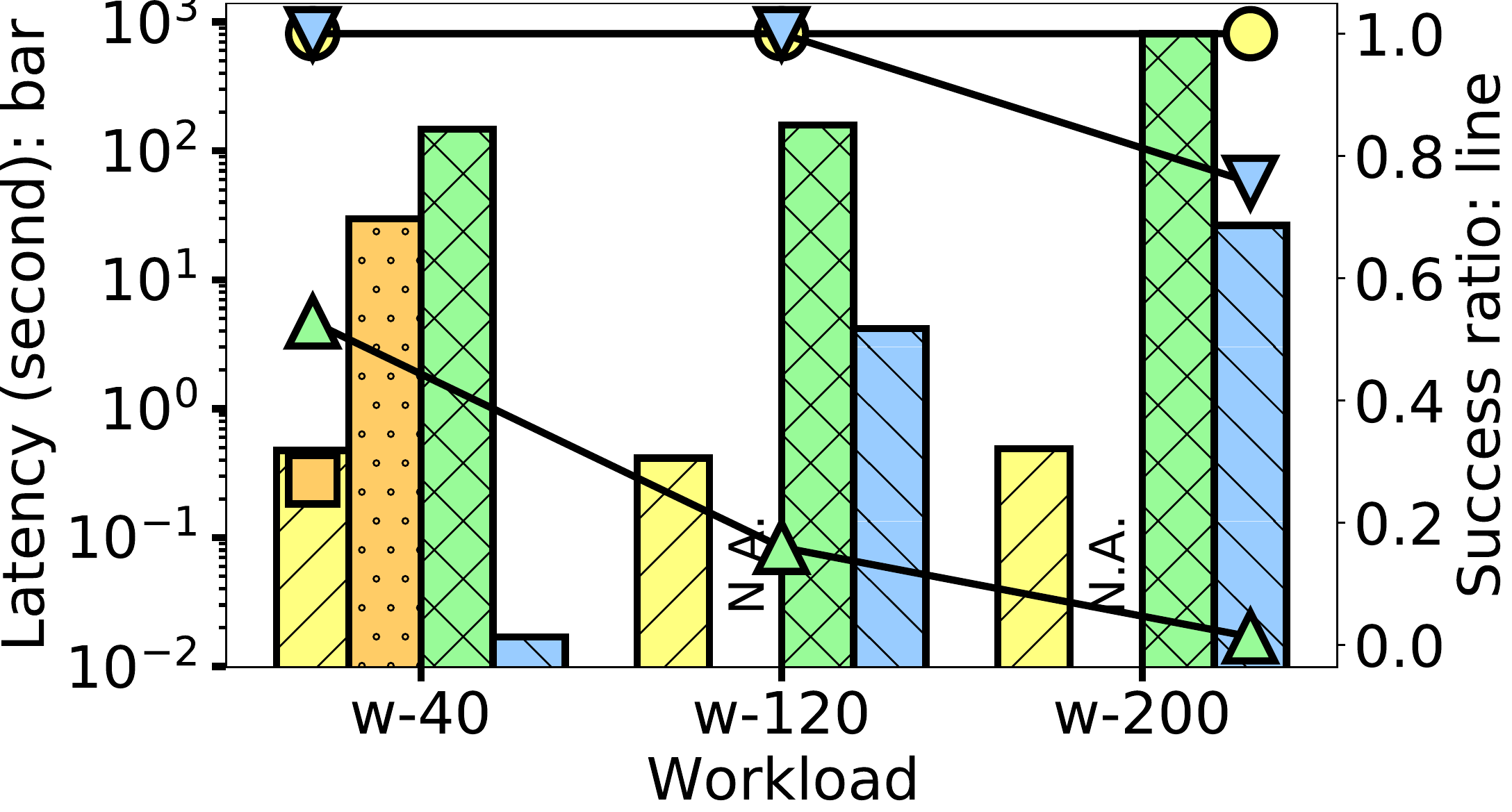}
    \vspace{-2mm}
    \caption{ALBERT (AWS)}
    \vspace{-1mm}
    \label{subfig:aws-albert}
\end{subfigure}
\vspace{1mm}

\begin{subfigure}[b]{0.48\columnwidth}
    \centering
    \includegraphics[width=0.98\columnwidth]{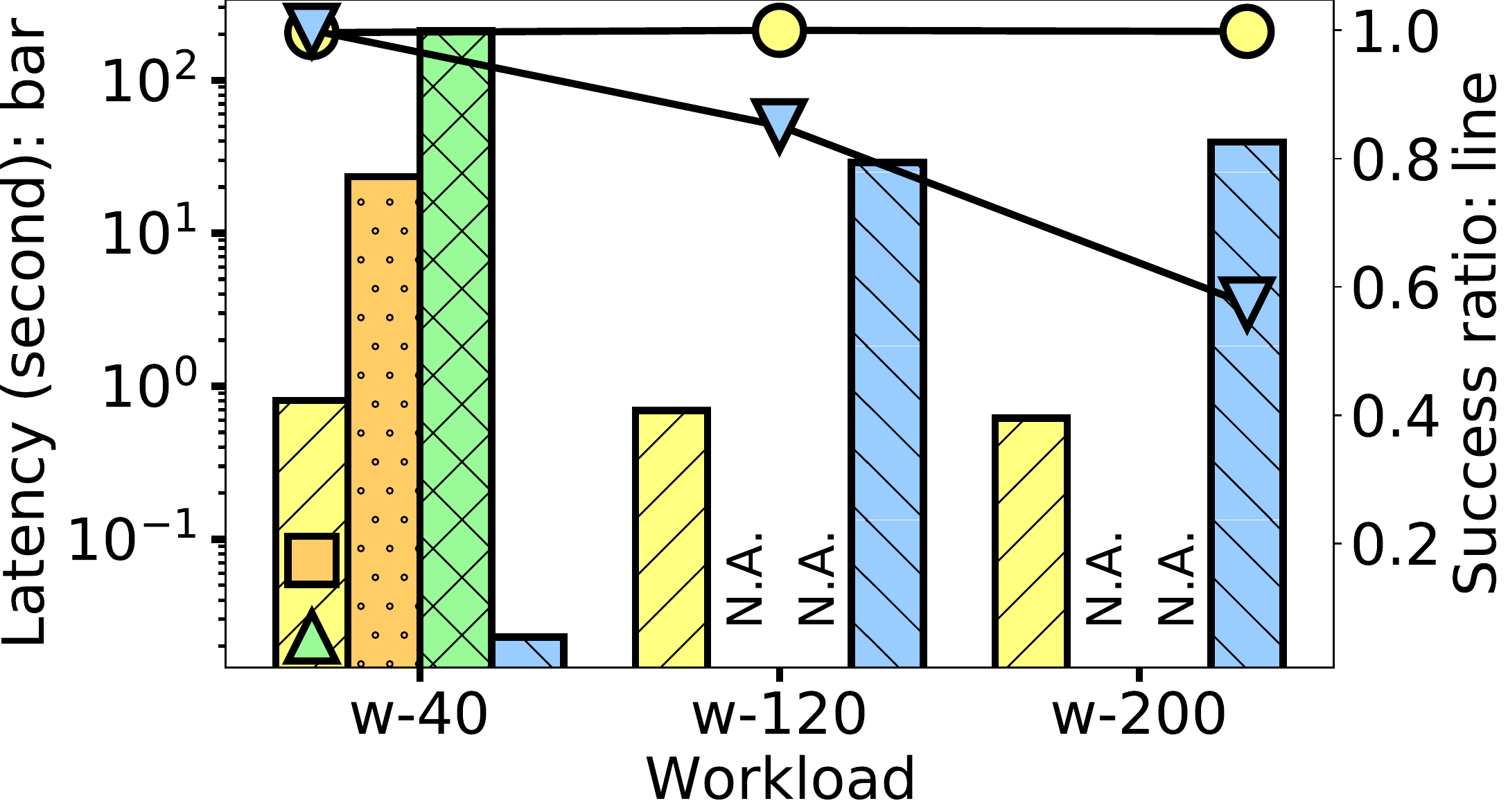}
    \vspace{-2mm}
    \caption{VGG (AWS)}
    \vspace{-1mm}
    \label{subfig:aws-vgg}
\end{subfigure}
~
\begin{subfigure}[b]{0.48\columnwidth}
    \centering
    \includegraphics[width=0.98\columnwidth]{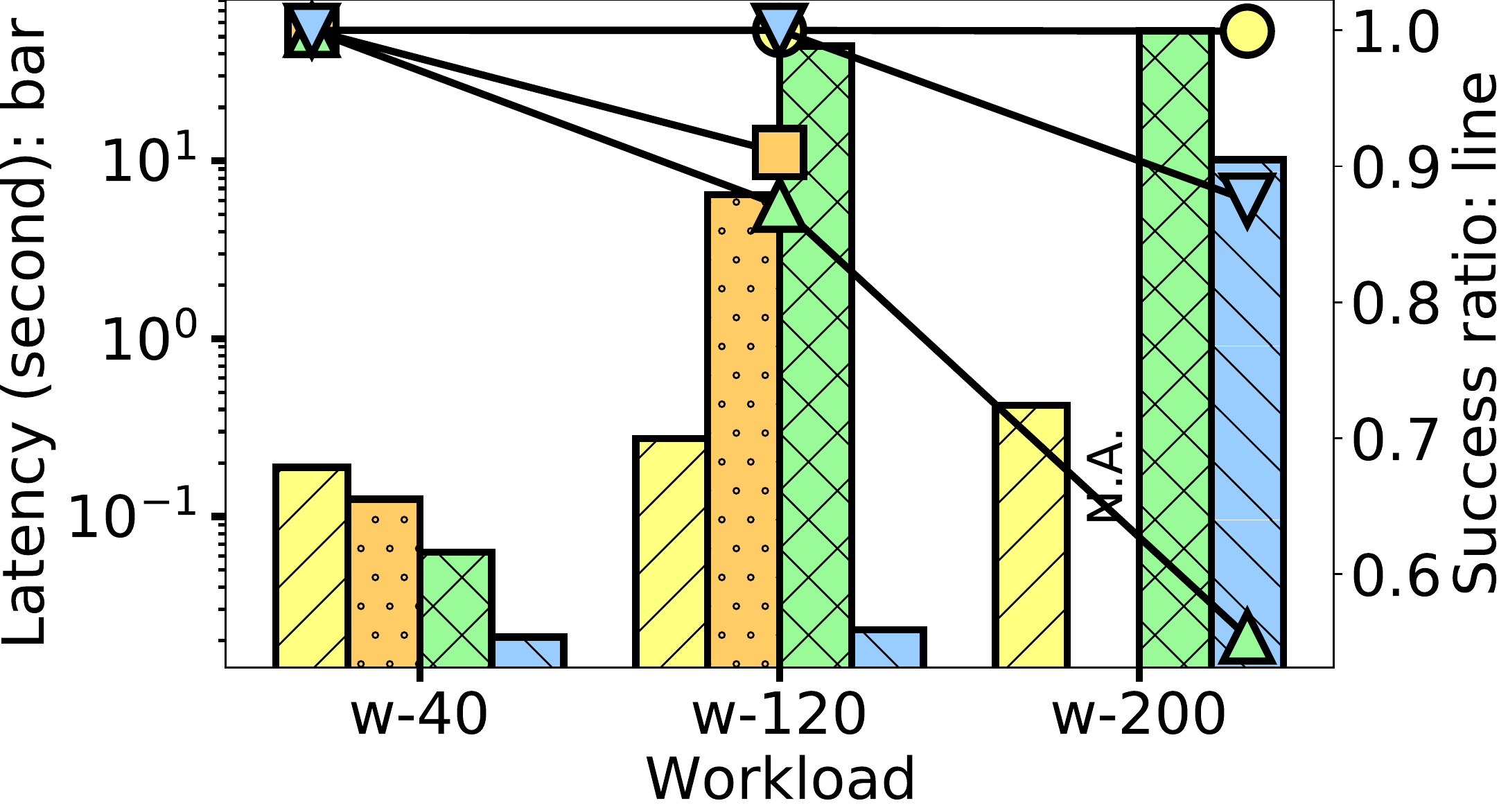}
    \vspace{-2mm}
    \caption{MobileNet (GCP)}
    \vspace{-1mm}
    \label{subfig:gcp-mobilenet}
\end{subfigure}
\vspace{1mm}

\begin{subfigure}[b]{0.48\columnwidth}
    \centering
    \includegraphics[width=0.98\columnwidth]{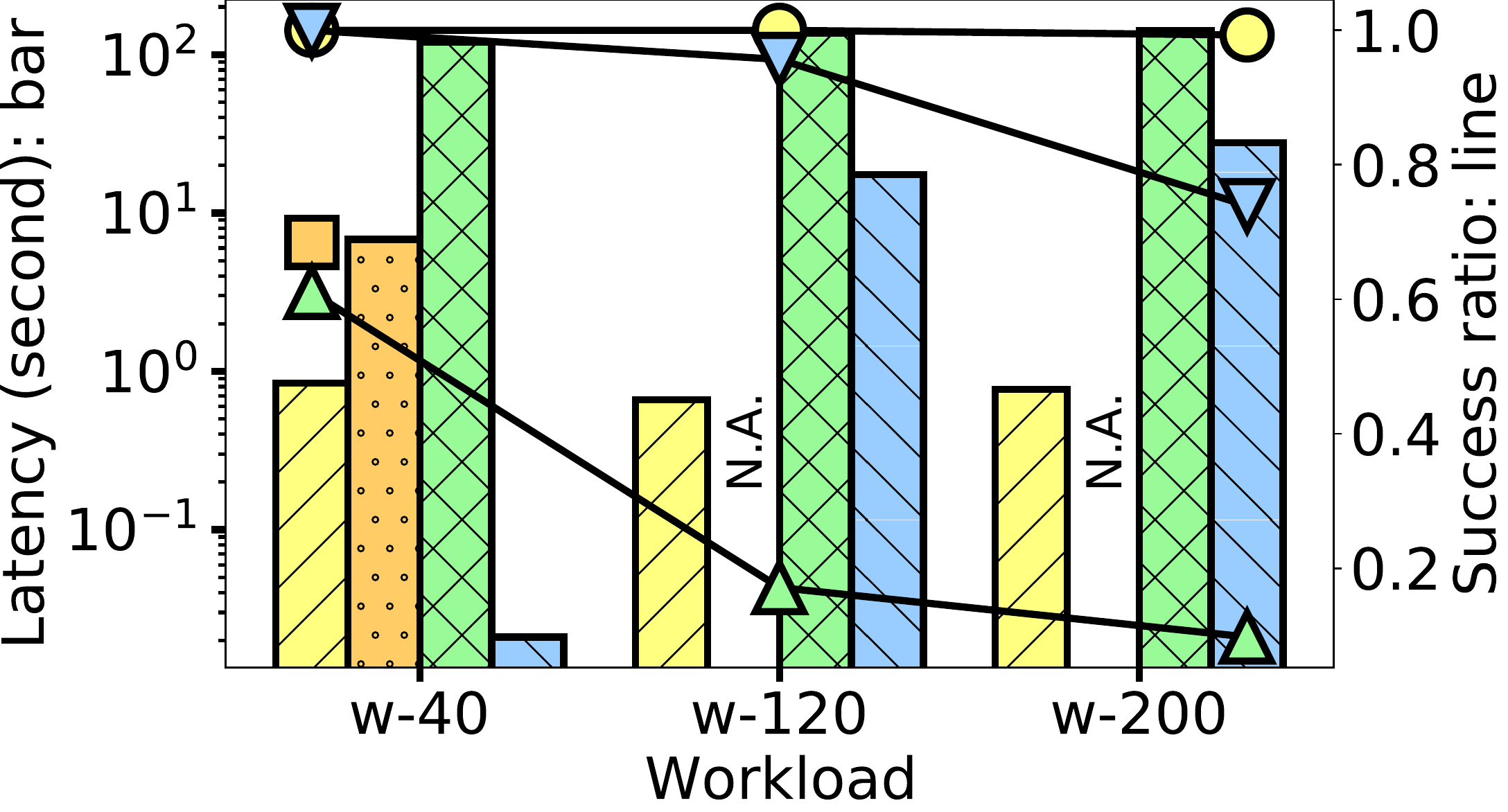}
    \vspace{-2mm}
    \caption{ALBERT (GCP)}
    \vspace{-1mm}
    \label{subfig:gcp-albert}
\end{subfigure}
~
\begin{subfigure}[b]{0.48\columnwidth}
    \centering
    \includegraphics[width=0.98\columnwidth]{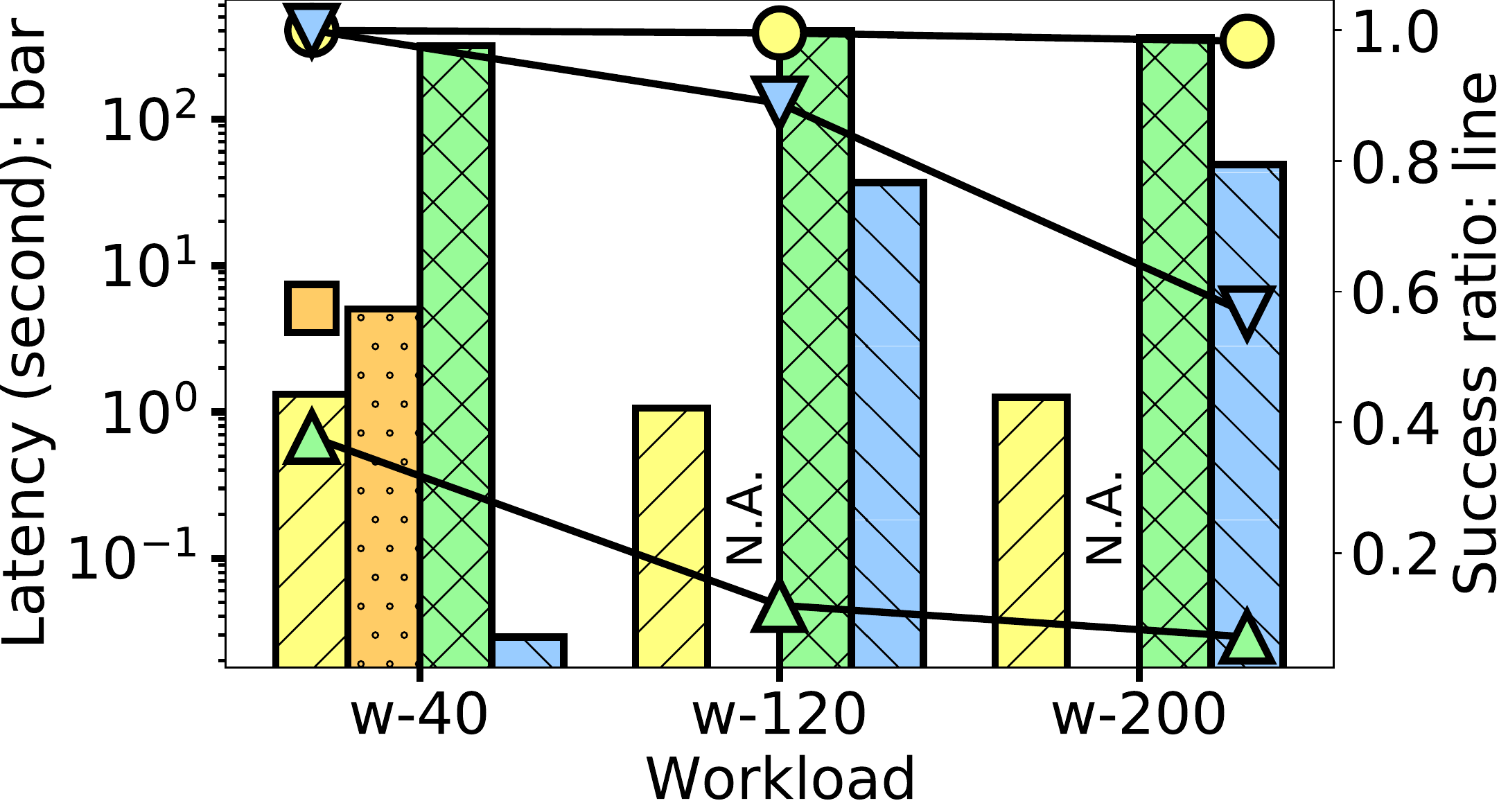}
    \vspace{-2mm}
    \caption{VGG (GCP)}
    \vspace{-1mm}
    \label{subfig:gcp-vgg}
\end{subfigure}
\vspace{-3mm}
\caption{Model serving systems' performance comparison.}
\vspace{-3mm}
\label{fig:serverless-vs-others}
\end{figure}
}

In this section, we compare the performance and cost of serverless model serving with alternative systems, including ManagedML, CPU server, and GPU server using serving runtime TensorFlow1.15. Figure~\ref{fig:serverless-vs-others} and Table~\ref{table:costs} summarize the overall comparison. 
%

\subsection{Summary of Key Findings}
\label{subsec:aws-google-summary}
We summarize three key findings before presenting the details. {\it First}, serverless has better performance and is more cost-effective than ManagedML in most cases. {Specifically, for MobileNet with workload-40, the average latency of AWS-ManagedML is 71.6$\times$ slower than that of AWS-Serverless, while its cost is 8.56$\times$ higher.} 
{\it Second}, in general, serverless outperforms CPU servers but often incurs a higher cost. However, on AWS, serverless can be better in both performance and cost when serving a simple model under a low workload. {For instance, for MobileNet with workload-40, AWS-Serverless is 104.5$\times$ faster than CPU server and also 1.78$\times$ lower in cost. }
{\it Third},  under a low workload, GPU servers are usually better than serverless. 
While under a high workload, serverless is more stable and could be a better choice than GPU servers. {For instance, for MobileNet with workload-200, AWS-Serverless results in an average latency of $0.097s$ and a cost of $\$0.186$, while those for a GPU server are $7.52s$ and $\$0.187$, respectively. In other words, AWS-Serverless is 77.5$\times$ faster given comparable cost.}

\begin{table}[t]
\centering
\scriptsize
\caption[]{Costs for evaluated model serving systems\footnotemark}
\vspace{-3mm}
\begin{tabular}{| c | c | c | c | c |}
\hline
\multicolumn{2}{|c|}{{Model Serving Systems}} & workload-40 & workload-120 & workload-200 \\
\hline
\hline
\multirow{3}{*}{{{AWS-Serverless}}} & {MobileNet} & {\$0.050} & {\$0.117} & {\$0.186} \\
& {ALBERT} & {\$0.223} & {\$0.665} & {\$1.326} \\
& {VGG} & {\$0.492} & {\$1.134} & {\$1.993} \\
\hline
\multirow{3}{*}{{AWS-ManagedML}} & MobileNet & \$0.428 & \$0.610 & - \\
& ALBERT & \$0.445 & - & - \\
& VGG & \$0.436 & - & - \\
\hline
\multicolumn{2}{|c|}{{AWS-CPU}} & \$0.089 & \$0.089 & \$0.092 \\
\multicolumn{2}{|c|}{{AWS-GPU}} & \$0.181 & \$0.182 & \$0.187 \\
\hline
\hline
\multirow{3}{*}{{{GCP-Serverless}}} & {MobileNet} & {\$0.065} & {\$0.279} & {\$0.537} \\
& {ALBERT} & {\$0.299} & {\$0.887} & {\$1.511} \\
& {VGG} & {\$0.507} & {\$1.438} & {\$2.467} \\
\hline
\multirow{3}{*}{{GCP-ManagedML}} & MobileNet & \$0.164 & \$0.313 & - \\
& ALBERT & \$0.468 & - & - \\
& VGG & \$0.872 & - & - \\
\hline
\multicolumn{2}{|c|}{{GCP-CPU}} & \$0.092 & \$0.092 & \$0.094 \\
\multicolumn{2}{|c|}{{GCP-GPU}} & \$0.176 & \$0.177 & \$0.182 \\
\hline
\end{tabular}
\label{table:costs}
\vspace{-3mm}
\end{table}

\subsection{Serverless vs. ManagedML Service} \label{subsec:compare-server-autoscaling}


We first compare serverless against managed ML services. For managed ML services, autoscaling is enabled, and the minimum number of running instances was set to 1. 

\noindent
\textbf{AWS-Serverless vs. AWS-ManagedML
} 
Figure~\ref{subfig:aws-mobilenet}-\ref{subfig:aws-vgg} show the comparison between AWS-Serverless and AWS-ManagedML for the MobileNet, ALBERT, and VGG models, respectively. The average latency of AWS-Serverless is two orders of magnitude lower than that of AWS-ManagedML. Furthermore, there are almost no failed requests in AWS-Serverless, but there are many in AWS-ManagedML, especially when the request rate is high or the model is complex. For example, the success ratio for MobileNet drops from 82\% (workload-40) to 36\% (workload-120). For ALBERT and VGG, even with workload-40, the success ratios are only 27\% and 17\%, respectively, rendering the service unusable.

\begin{figure}[t]
\hspace{-3mm}
\includegraphics[width=0.99\columnwidth]{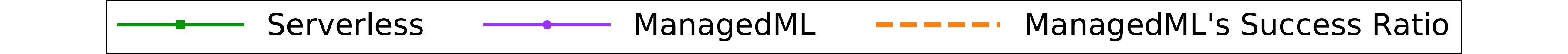}
\begin{subfigure}[b]{0.23\textwidth}
    \centering
    \includegraphics[width=0.98\columnwidth]{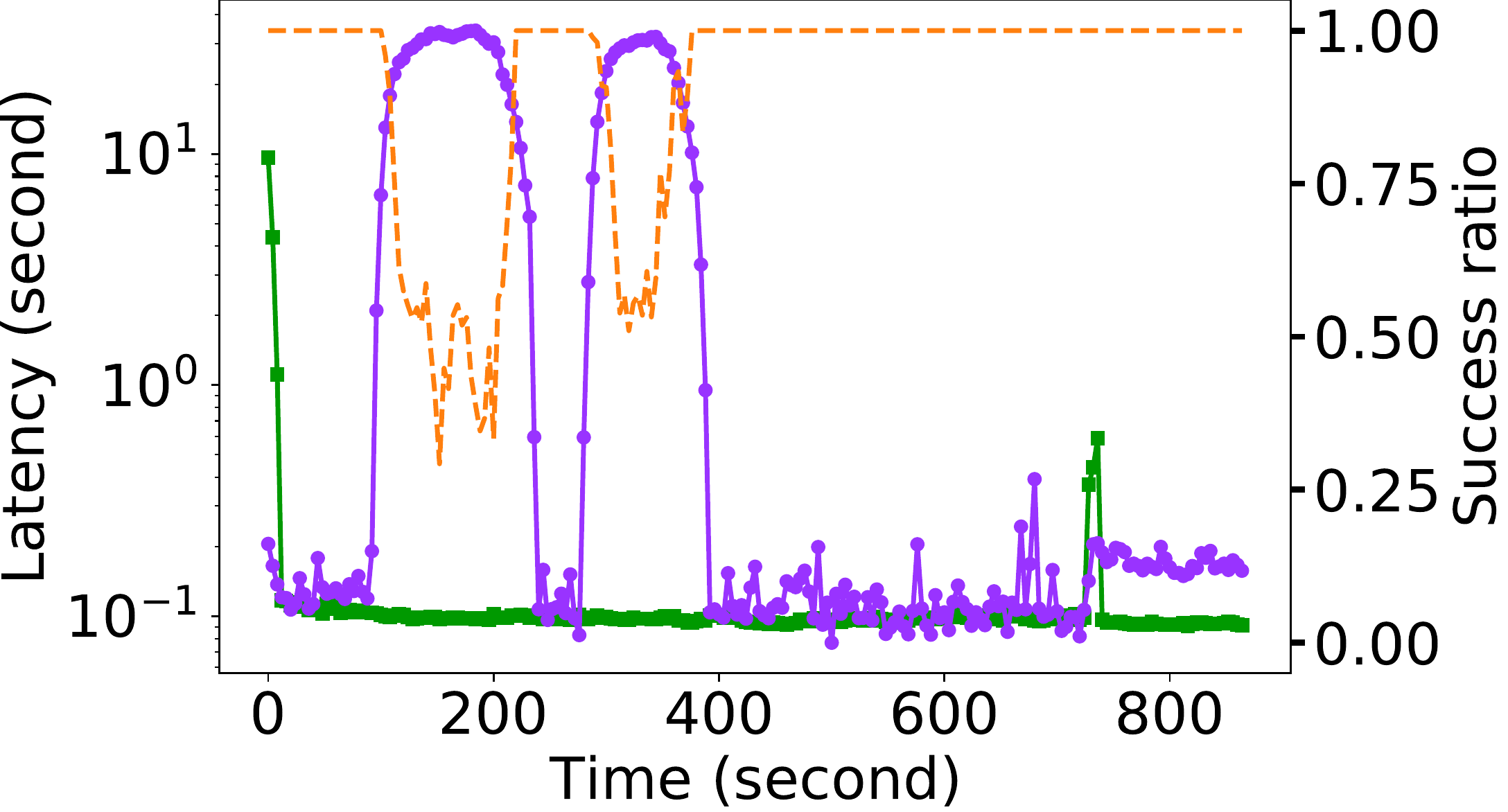}
    \vspace{-2mm}
    \caption{MobileNet with w-40 (AWS)}
    \label{subfig:lambda-sagemaker-mb40}
\end{subfigure}
~
\begin{subfigure}[b]{0.23\textwidth}
    \centering
    \includegraphics[width=0.98\columnwidth]{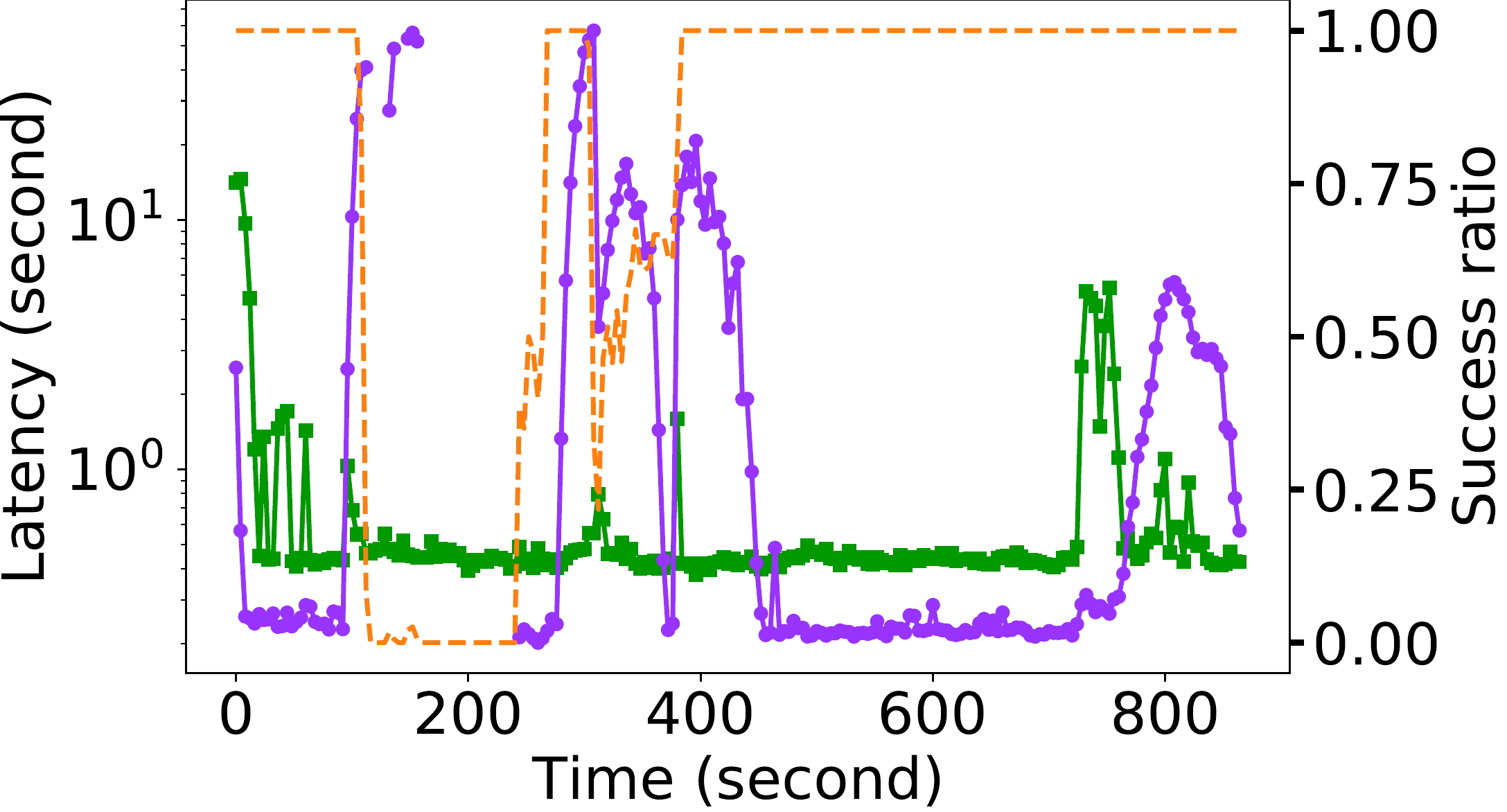}
    \vspace{-2mm}
    \caption{ALBERT with w-40 (GCP)}
    \label{subfig:cf-ai-ab40}
\end{subfigure}
\vspace{-4mm}
\caption{Serverless and ManagedML comparison.}
\vspace{-3mm}
\label{fig:serverless-vs-mlservice}
\end{figure}

\begin{figure}[t]
\centering
\begin{subfigure}[b]{0.21\textwidth}
    \centering
    \includegraphics[width=0.98\columnwidth]{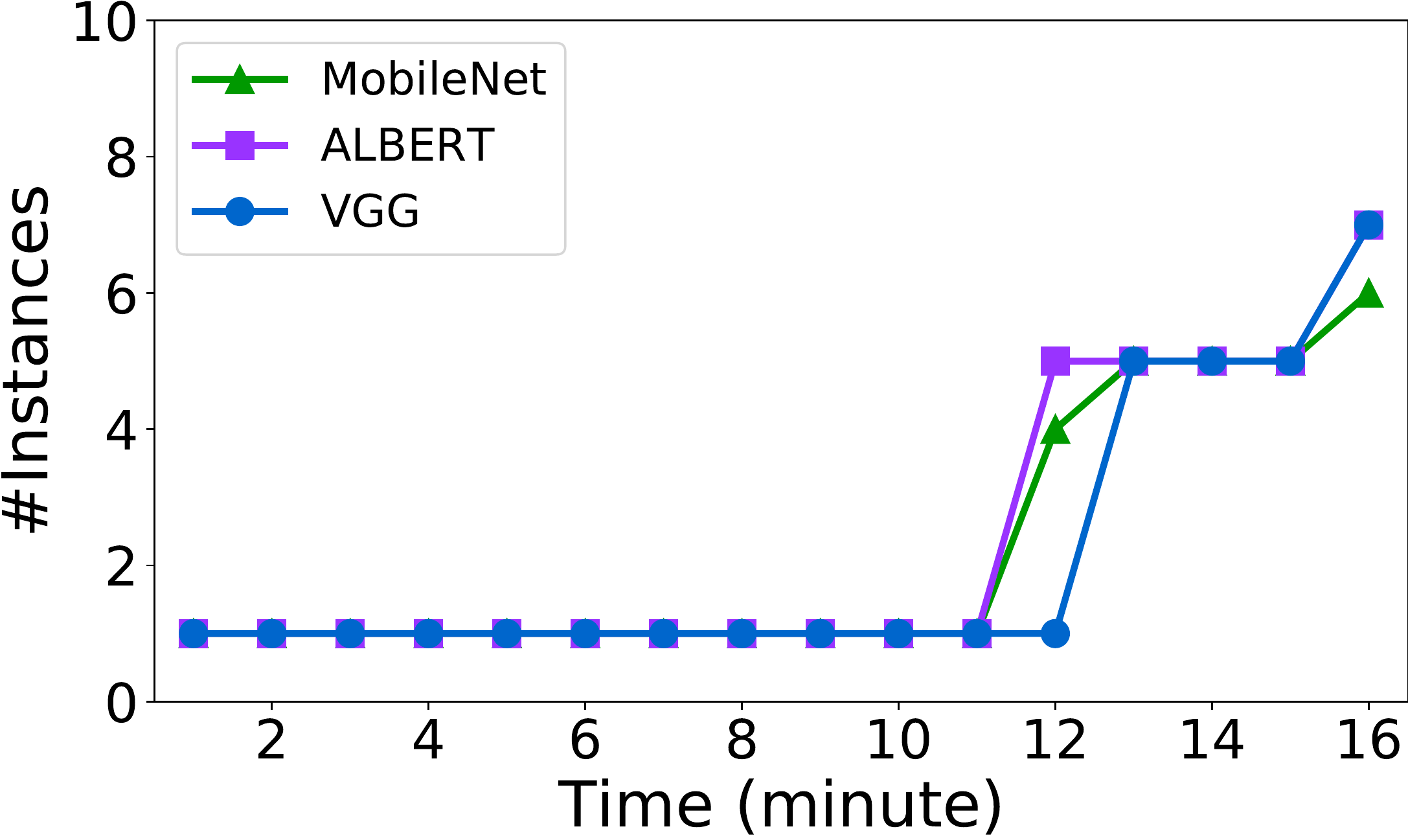}
    \vspace{-2mm}
    \caption{AWS ManagedML}
    \label{subfig:aws-ml-w40-instance-num}
\end{subfigure}
~
\begin{subfigure}[b]{0.21\textwidth}
    \centering
    \includegraphics[width=0.98\columnwidth]{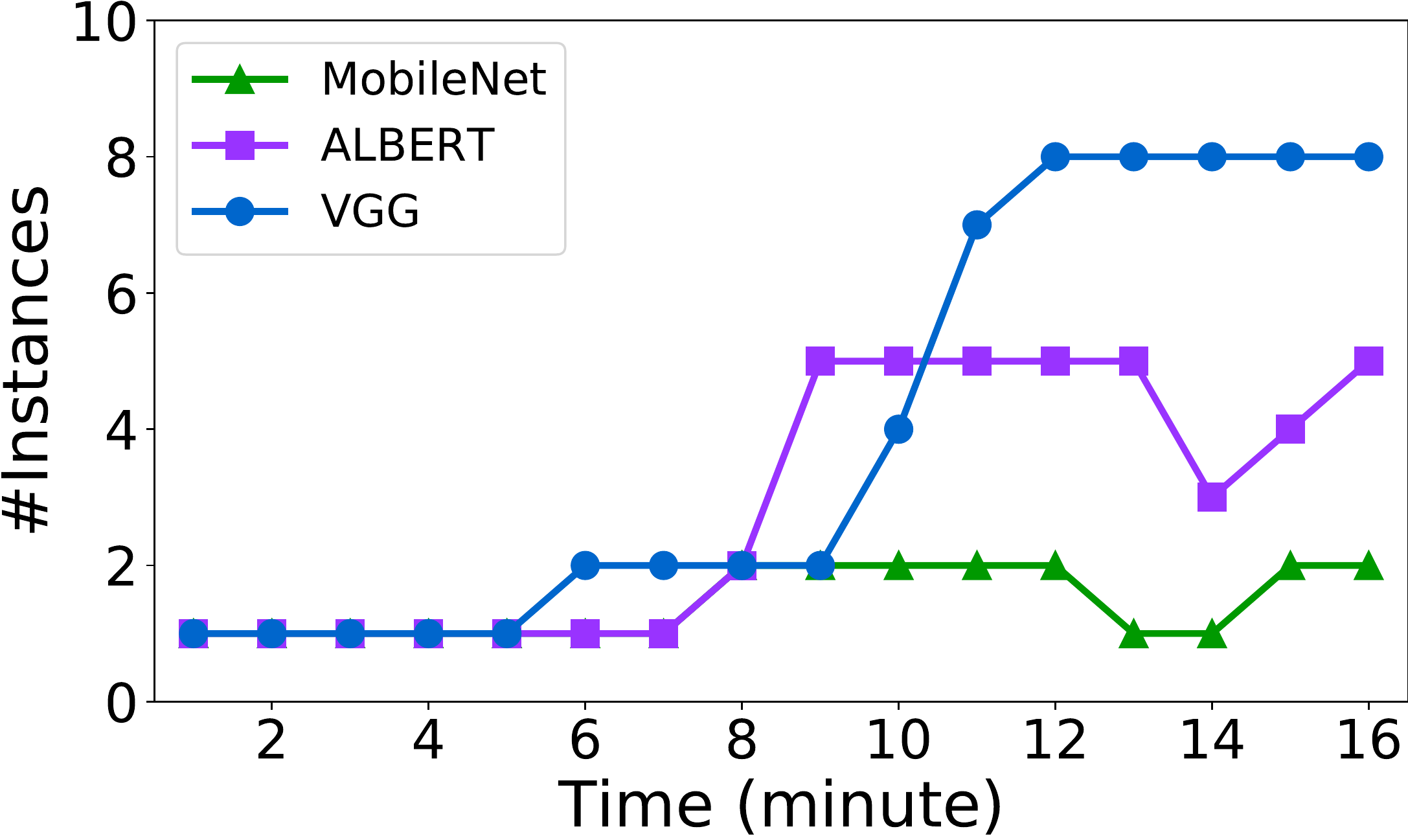}
    \vspace{-2mm}
    \caption{GCP ManagedML}
    \label{subfig:gcp-ml-w40-instance-num}
\end{subfigure}
\vspace{-4mm}
\caption{The number of instances on ManagedML services.}
\vspace{-3mm}
\label{fig:ml-service-instance-num}
\end{figure}

{Figure~\ref{subfig:lambda-sagemaker-mb40} shows a detailed comparison for MobileNet under workload-40, where solid lines denote latency and dotted lines denote success ratio.} At first, the average latency of AWS-Serverless (about 10 seconds) is higher than that of AWS-ManagedML. This is due to the cold-start time. 
However, as the system is warming up
, AWS-Serverless performs better and is more stable than AWS-ManagedML. When the request rate becomes high (e.g., starting at around timestamp 100), AWS-ManagedML is unable to keep up, resulting in high latency and request failure.  
%
%
In fact, AWS-ManagedML's autoscaling takes several minutes to start new instances, leading to a large number of queued requests and thus delayed responses. 
{
Figure~\ref{subfig:aws-ml-w40-instance-num} shows the number of active instances on AWS-ManagedML under workload-40. We can observe on the platform that it desires about 5 instances at timestamp 7 minutes. However, the instances are ready and in service at timestamp 11 minutes.
While as will be shown in Figure~\ref{subfig:lambda-sagemaker-w40-instance-num}, AWS-Serverless can spawn new instances (e.g., tens or even a hundred of instances) quickly under the same workload for its high elasticity. 
}
%
%


Regarding cost from Table~\ref{table:costs}, we can see that AWS-Serverless is more cost-efficient than AWS-ManagedML. The reason is that AWS-ManagedML needs several minutes to start new instances, and the evaluated workloads are relatively short. Most of the costs are spent on autoscaling instances rather than on doing the prediction. 

\footnotetext{The costs are the absolute values for the evaluated experiments.}

\noindent
\textbf{GCP-Serverless vs. GCP-ManagedML. }
Figure~\ref{subfig:gcp-mobilenet}-\ref{subfig:gcp-vgg} show the results for Google systems under the same settings. 
For MobileNet under workload-40, there are no failed requests for both systems, but GCP-Serverless is slightly worse than GCP-ManagedML regarding latency, which is different from the comparison on AWS. The reason may be two-fold. One is that GCP-Serverless performs relatively worse than AWS-Serverless, which will be explored in detail in Section~\ref{subsec:serverless-comparison}. 
The other is that GCP-ManagedML scales slightly better than AWS-ManagedML in the experiments. 
As shown in Figure~\ref{subfig:gcp-ml-w40-instance-num},
GCP-ManagedML can scale to 2 instances at timestamp 6 minutes, which is earlier than AWS-ManagedML. Therefore, GCP-ManagedML can process the requests relatively faster, which results in better performance. 
%
%
%
However, when serving a larger model or with a higher workload, both latency and success ratio of GCP-ManagedML deteriorate significantly, while those of GCP-Serverless are stable and better.
Figure~\ref{subfig:cf-ai-ab40} details the comparison for ALBERT with workload-40. 
Similarly, 
once the number of queued requests reaches a threshold, its performance degrades quickly. 

%

%
For the cost, GCP-Serverless is slightly more efficient, but the advantage is not significant as the comparison on AWS. This is mainly because GCP-Serverless does not perform as well as AWS-Serverless. Nevertheless, if considering both performance and cost, GCP-Serverless is still a preferable choice over GCP-ManagedML.

\subsection{Serverless vs. CPU Server}\label{subsec:compare-cpu}

\begin{figure}[t]
\hspace{-3mm}
\includegraphics[width=0.99\columnwidth]{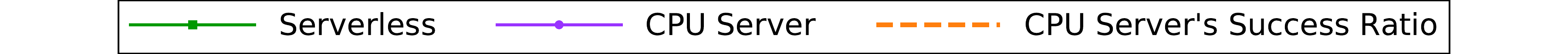}
\begin{subfigure}[b]{0.23\textwidth}
    \centering
    \includegraphics[width=0.98\columnwidth]{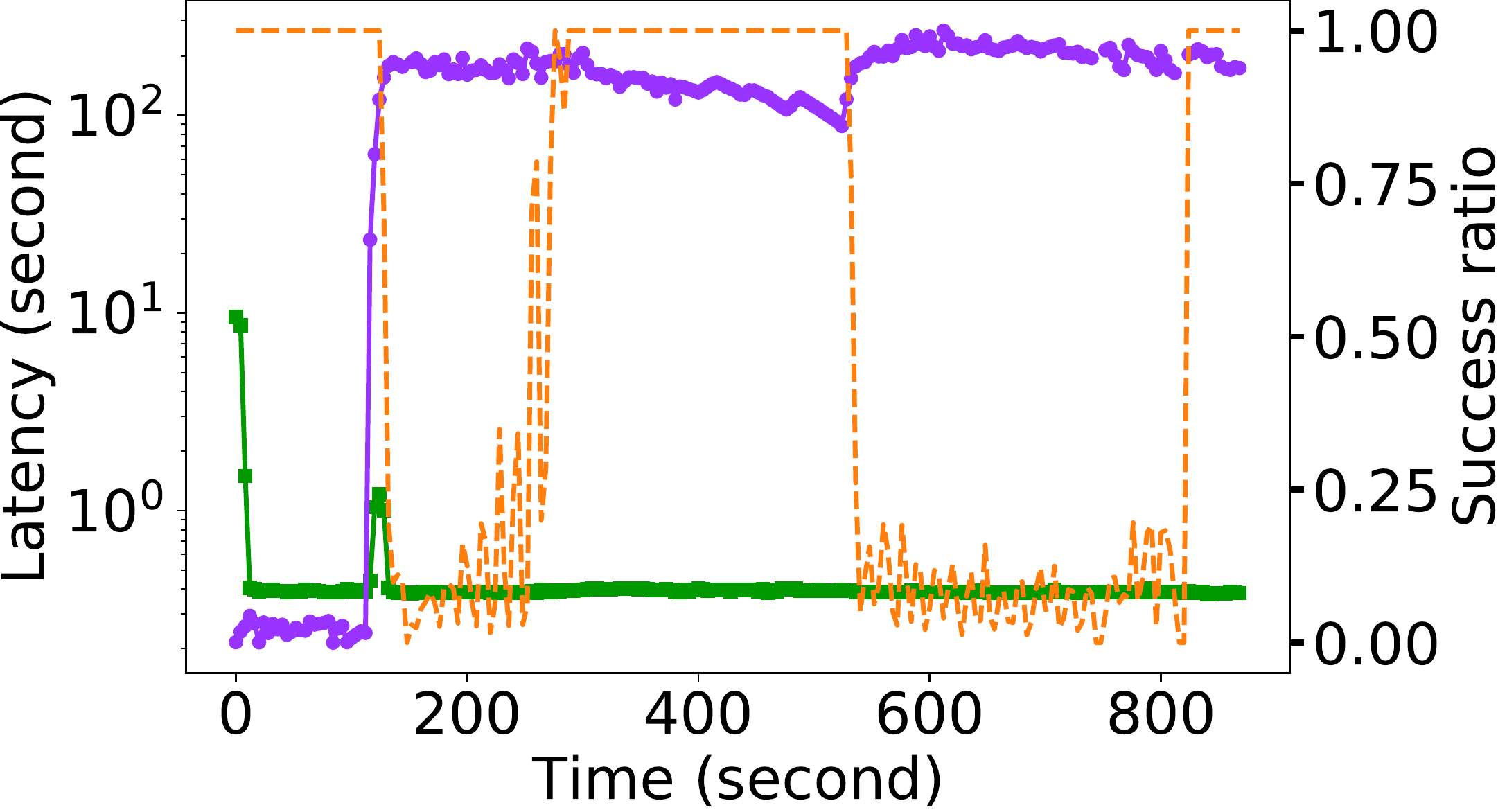}
    \vspace{-2mm}
    \caption{ALBERT with w-120 (AWS)}
    \label{subfig:lambda-cpu-mb120}
\end{subfigure}
~
\begin{subfigure}[b]{0.23\textwidth}
    \centering
    \includegraphics[width=0.98\columnwidth]{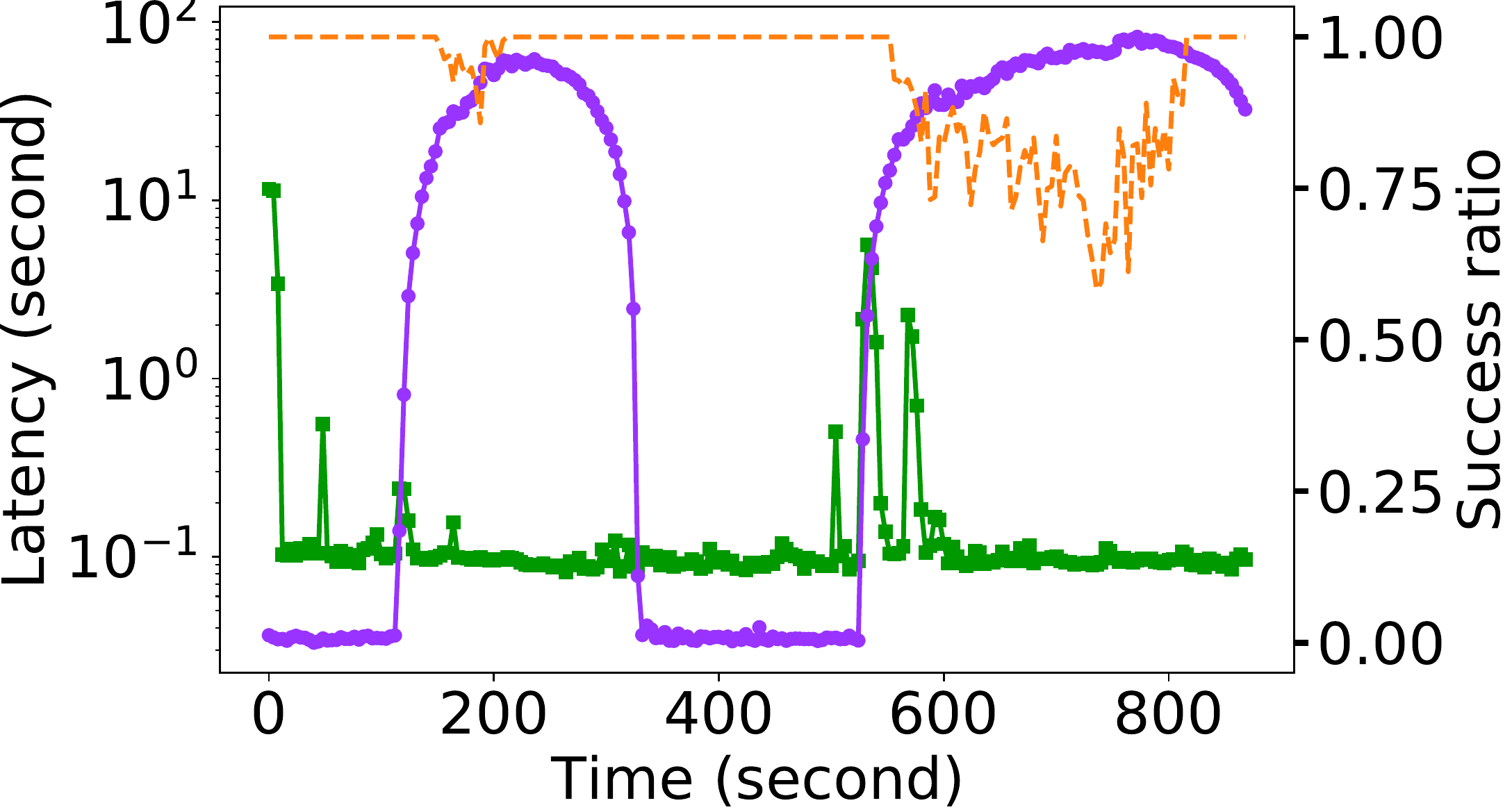}
    \vspace{-2mm}
    \caption{MobileNet with w-120 (GCP)}
    \label{subfig:cf-cpu-mb120}
\end{subfigure}
\vspace{-4mm}
\caption{Serverless and CPU server comparison.}
\vspace{-3mm}
\label{fig:serverless-vs-cpu}
\end{figure}

\noindent
\textbf{AWS-Serverless vs. CPU server.} 
From Figure~\ref{subfig:aws-mobilenet}-\ref{subfig:aws-vgg}, we observe that for all models, the average latency of AWS-Serverless is always smaller than CPU server, and the advantage is more pronounced when the workload is higher or when the model is larger. 
%
For example, for the CPU server, the success ratios of MobileNet are 100\%, 44\%, and 27\% with workloads 40, 120, and 200, respectively; meanwhile, under workload-40, the success ratios for MobileNet, ALBERT, and VGG are 100\%, 53\%, and 6\%, respectively. 
This is because the CPU server is overloaded such that more requests are queued up under higher load or the execution time per request is longer, leading to increased latency and failure requests. 
Figure~\ref{subfig:lambda-cpu-mb120} details the performance of ALBERT with workload-120, from which we see the latency goes up sharply at the first request peak (timestamp 100) and stays at a high level. In contrast, AWS-Serverless' latency remains consistently low for its superior elasticity. 

AWS EC2 also provides autoscaling for self-rented servers; thus, we create an autoscaling group for each serving model, such that new instances will be started and added in the group given a pre-defined template. Then, we create a load balancer for the group to receive clients' requests and forward the requests to existing instances. Nonetheless, similar to what we observed on AWS-ManagedML, it takes several minutes (i.e., 3 to 5 minutes) to start a new instance, making the service less reactive to bursty requests. Besides, there was no direct way to collect the cost of autoscaled instances on AWS EC2; thus, we do not report the results in this paper.

For the cost, AWS-Serverless is higher in most cases. However, we note that the average latency and success ratio of the CPU server are extremely low. A comparable situation is for MobileNet with workload-40, where AWS-Serverless is cheaper (i.e., \$0.065) than CPU server (i.e., \$0.089) while delivering better performance.

\noindent
\textbf{GCP-Serverless vs. CPU server.}   Figure~\ref{subfig:gcp-mobilenet}-\ref{subfig:gcp-vgg} illustrate a similar pattern to the comparison between GCP-Serverless and GCP-ManagedML.
%
In particular, for MobileNet with workload-40, CPU server is slightly faster than GCP-Serverless. However, when the workload increases, the performance of CPU server degrades greatly, as shown in Figure~\ref{subfig:cf-cpu-mb120}. At the two request peaks (timestamps 100 to 250 and 500 to 800), the average latency grows tens of seconds.
%
For the cost, GCP-Serverless is more expensive. However, as mentioned above, CPU server cannot handle bursty requests effectively.

\subsection{Serverless vs. GPU Server}\label{subsec:compare-gpu}

\begin{figure}[t]
\includegraphics[width=0.98\columnwidth]{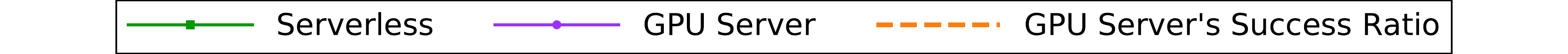}
\begin{subfigure}[b]{0.23\textwidth}
    \centering
    \includegraphics[width=0.98\columnwidth]{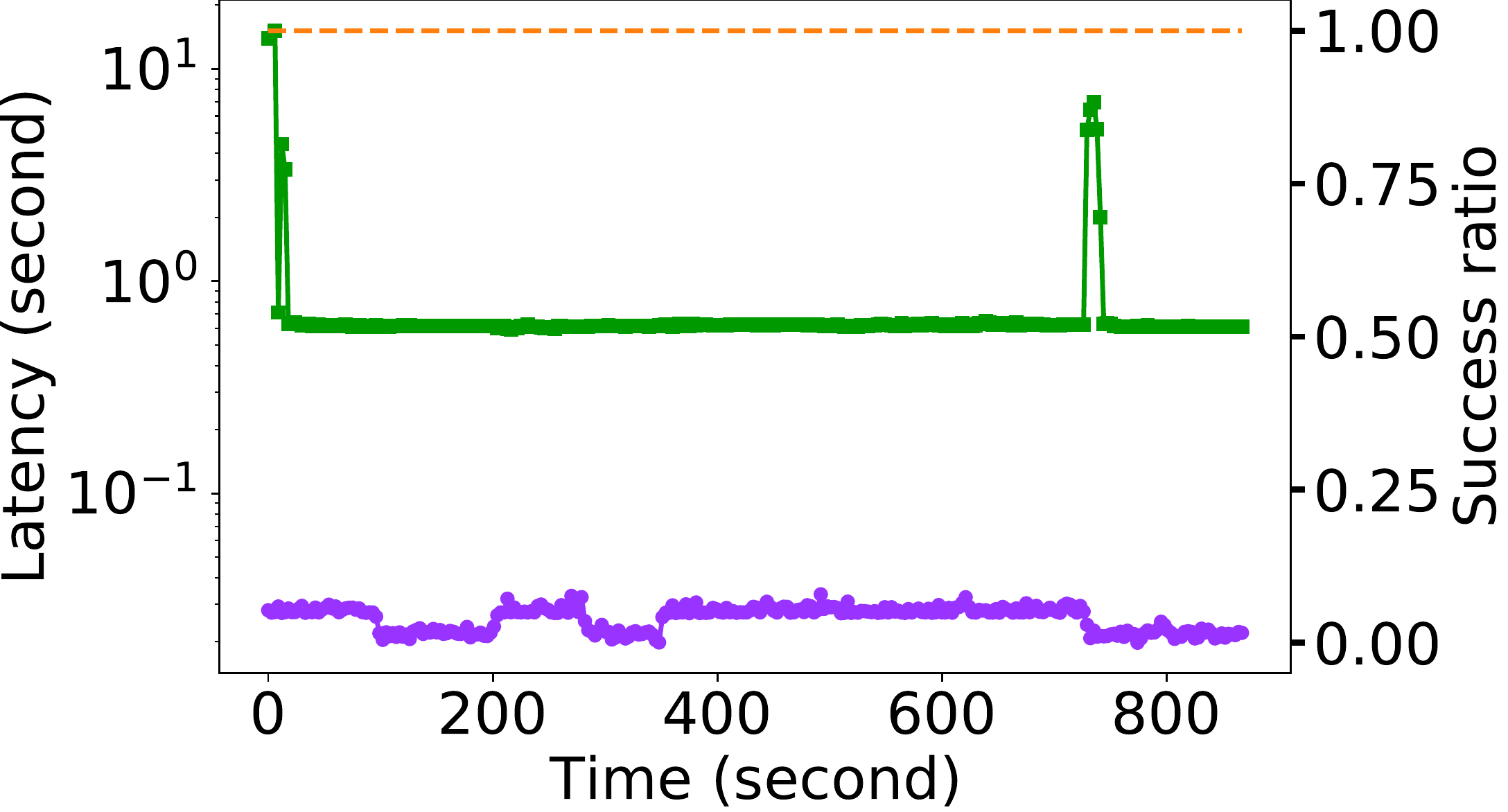}
    \vspace{-2mm}
    \caption{VGG with w-40 (AWS)}
    \label{subfig:lambda-gpu-vgg40}
\end{subfigure}
~
\begin{subfigure}[b]{0.23\textwidth}
    \centering
    \includegraphics[width=0.98\columnwidth]{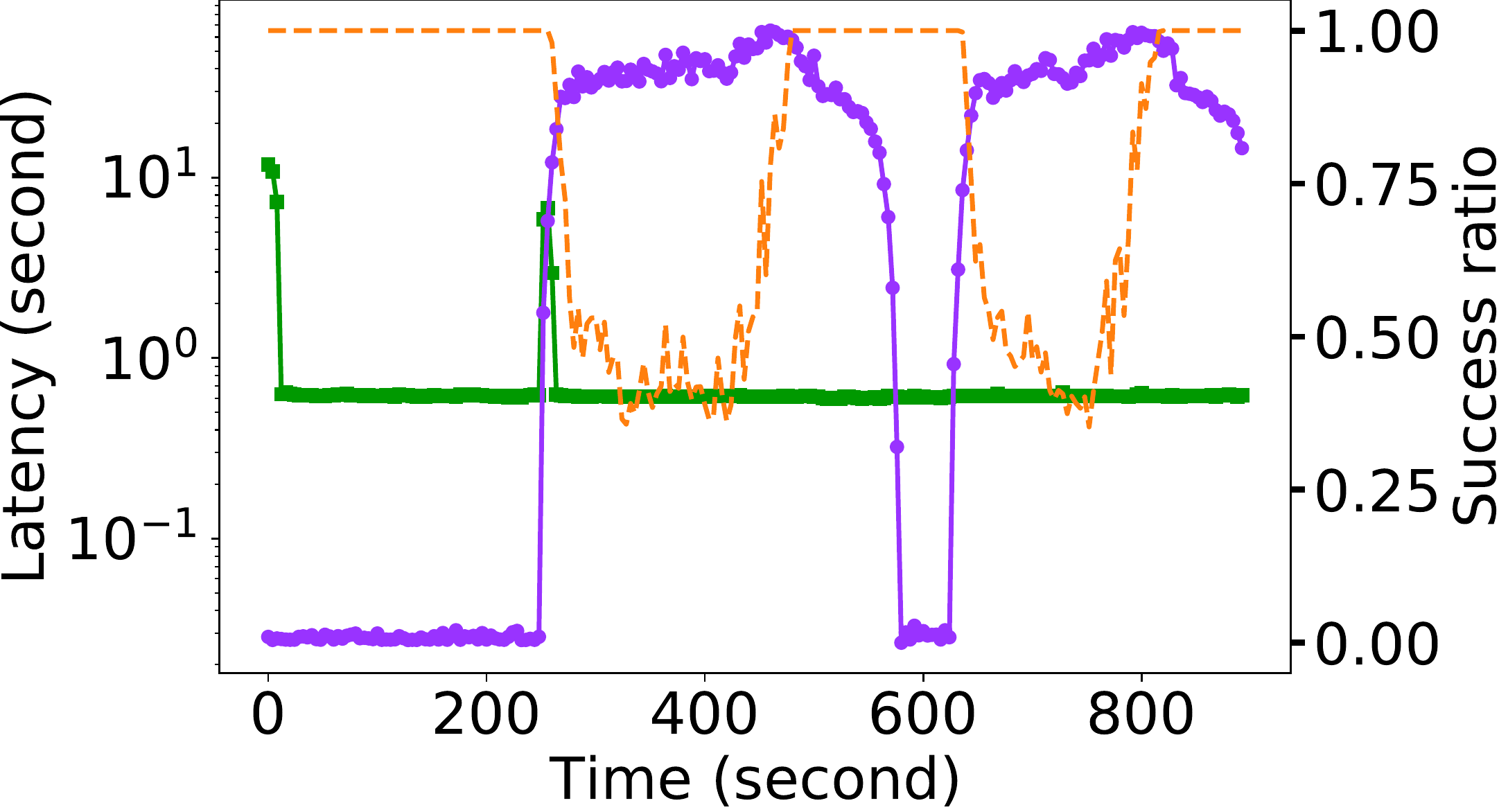}
    \vspace{-2mm}
    \caption{VGG with w-200 (AWS)}
    \label{subfig:lambda-gpu-vgg200}
\end{subfigure}

\vspace{-4mm}
\caption{Serverless and GPU server comparison.}
\vspace{-3mm}
\label{fig:serverless-vs-gpu}
\end{figure}

\noindent
\textbf{AWS-Serverless vs. GPU server.}
Figure~\ref{subfig:aws-mobilenet}-\ref{subfig:aws-vgg} show that under workload-40, the GPU server's average latency is always lower than that of AWS-Serverless for the three models. Figure~\ref{subfig:lambda-gpu-vgg40} gives a detailed comparison for the VGG model. That is because the GPU server can process each request quickly (e.g., about 0.02 seconds per request in our experiments). 
%
%
%
However, when given a higher workload, more requests are queued up, as shown in Figure~\ref{subfig:lambda-gpu-vgg200} for VGG with workload-200. The results can be analyzed in three stages.
First, at the beginning, the GPU server performs better than AWS-Serverless since the latter incurs cold-start overhead.  
Second, once AWS-Serverless instances are warmed up, AWS-Serverless outperforms the GPU server when the request rates are high. The rationale is that the request rate exceeds GPU server's capacity; thus, the request queue grows and leads to higher latency.  
Third, when the request rates are reduced, GPU server regains its advantage (e.g., around timestamp 600). However, in most periods, AWS-Serverless' latency is less than GPU server, resulting in lower average latency.

Regarding the cost, there are two main observations. First, for the MobileNet model, GPU server has better performance under workload-40 and workload-120, but its cost is also higher because the GPU server is under-utilized while still being charged. 
While for large models, e.g., ALBERT and VGG, GPU server is cheaper. The reason is that AWS-Serverless is charged by the number of requests and duration for processing each request. The execution time for these two models is relatively expensive, leading to a higher cost. 
Second, under a low workload, GPU server is better 
in both cost and performance in most cases,  
demonstrating its superior ability for model serving. However, under a high workload, GPU server has higher latency and request failures, and AWS-Serverless can be a better choice as its latency is less sensitive to the workloads and models, demonstrating its high elasticity. {For example, for MobileNet with workload-200, AWS-Serverless is 77.5$\times$ faster than GPU Server given comparable cost.}
In addition, as mentioned in Section~\ref{subsec:compare-cpu}, it does not perform well to use autoscaling on CPU servers. More severely, we observe that when using autoscaling on GPU servers, it takes 8 to 10 minutes before a new instance can serve requests, rendering autoscaling useless in our workloads. 


\noindent
\textbf{GCP-Serverless vs. GPU server.}
The comparison 
is similar to that on AWS. We omit the description due to the space limitation. 
\section{Design Space of Serverless Serving} \label{sec:investigation}

So far, we have shown that serverless is a viable and promising option for model serving.
%
In this section, we further investigate the impact of serverless serving's design space on its performance, including serverless platforms, serving runtimes, and various function-specific parameters. 

\subsection{Serverless Platforms}
\label{subsec:serverless-comparison}


We first compare the performance and cost of serverless platforms, as shown in Figure~\ref{fig:serverless-vs-others} and Table~\ref{table:costs}. We observe that AWS-Serverless is always better for the three metrics. {For instance, given MobileNet under workload-200 (see Figure~\ref{subfig:aws-vgg} and Figure~\ref{subfig:gcp-vgg}), the average latency, success ratio, and cost on AWS-Serverless are 0.097{\it s}, 100\%, and \$0.186, while those on GCP-Serverless are 0.422{\it s}, 99.94\%, and \$0.537, respectively.} There are two main reasons.

\begin{figure}[t]
    \centering
    \includegraphics[width=0.4\textwidth]{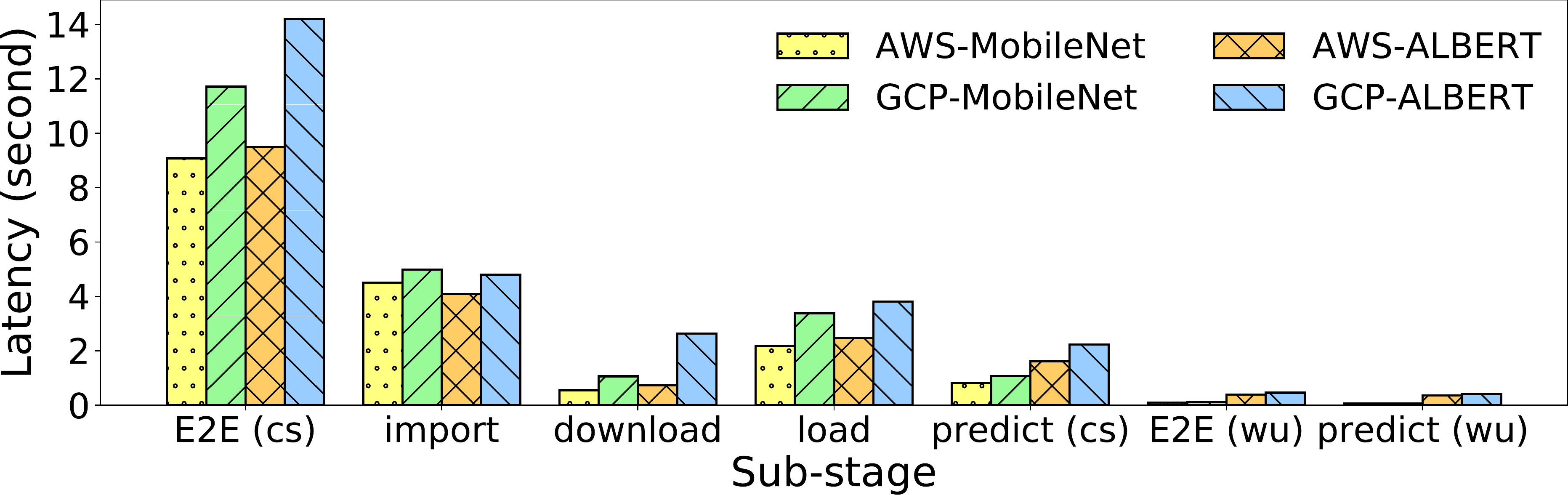}
    \vspace{-4mm}
\caption{Breakdown comparison of serverless platforms, where "cs" and "wu" denote "cold-start" and "warm-up".}
\vspace{-3mm}
\label{fig:break-down-comp}
\end{figure}

{
{First}, one contributing factor is the instance's cold-start time. In particular, GCP-Serverless takes a longer time to execute the cold-start requests, for example, around 11.71$s$ and 14.19$s$ for MobileNet and ALBERT under workload-120, respectively. In contrast, AWS-Serverless takes only 9.08$s$ and 9.49$s$, respectively. 
To attribute the difference, we break down the latency into cold-start and warm-up requests with sub-stages. For cold-start requests, we report: {\it end-to-end} (i.e., E2E) latency that between sending a request and receiving the result on each client; {\it import} time for importing the serving dependencies (e.g., TF1.15), {\it download} time for downloading the model (e.g., MobileNet) from cloud storage, {\it load} time for loading the model into the serving runtime, and {\it predict} time for calculating the model inference. For warm-up requests, we only report the E2E latency and predict time as the other sub-stages are not executed.

Figure~\ref{fig:break-down-comp} shows the breakdown comparison between the two serverless platforms given MobileNet and ALBERT with workload-120. We observe that the import sub-stage takes around 4-5 seconds and dominates the E2E cold-start latency for the two models on both platforms. 
Meanwhile, AWS-Serverless is faster than GCP-Serverless for all sub-stages (e.g., 1.89$s$ and 1.34$s$ faster for ALBERT {\it w.r.t.} the download time and load time, respectively), which demonstrates the AWS-Serverless' superior performance.
%
Besides, the predict time of cold-start requests is much longer than that of warm-up requests on both platforms. This may be because the TensorFlow runtime has components that are lazily initialized, which can cause a higher latency for the first request sent to a loaded model~\cite{tensorflow-saved-model}.
}
%

\begin{figure}[t]
\centering
\begin{subfigure}[b]{0.21\textwidth}
    \centering
    \includegraphics[width=0.98\columnwidth]{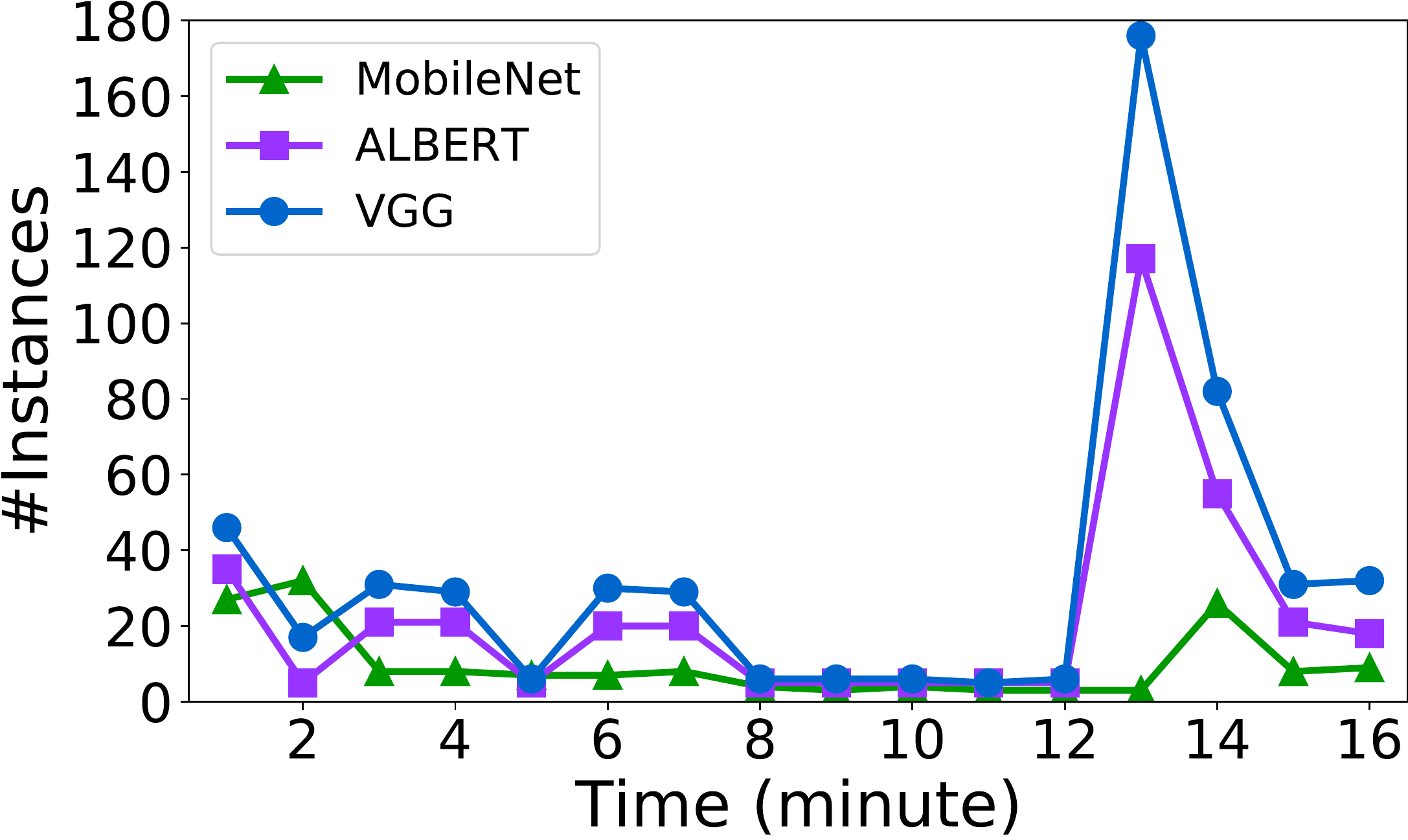}
    \vspace{-2mm}
    \caption{AWS Serverless}
    \label{subfig:lambda-sagemaker-w40-instance-num}
\end{subfigure}
~
\begin{subfigure}[b]{0.21\textwidth}
    \centering
    \includegraphics[width=0.98\columnwidth]{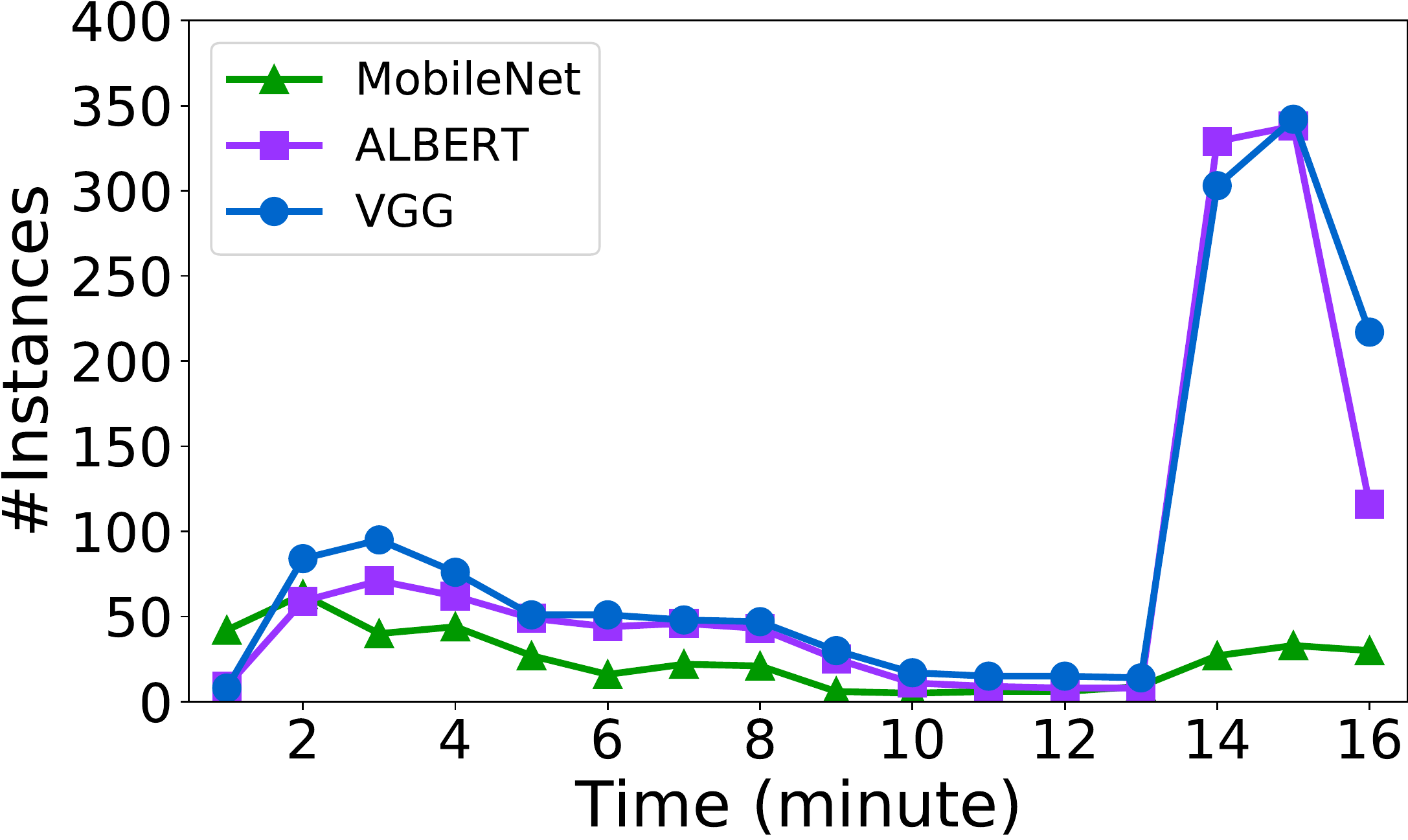}
    \vspace{-2mm}
    \caption{GCP Serverless}
    \label{subfig:cf-ai-w40-instance-num}
\end{subfigure}
\vspace{-4mm}
\caption{The number of instances on serverless platforms.}
\vspace{-3mm}
\label{fig:instance-num}
\end{figure}

{Second}, the long cold-start time also leads to over-provisioning problem~\cite{mark}, i.e., more instances are created than needed, as the service is still insufficient during the instance creation process, so that serverless platforms keep starting instances until the service is available. Figure~\ref{fig:instance-num} compares the number of active instances for the three models under workload-40. 
{We observe that both platforms can scale very fast (i.e., up to hundreds of instances in one minute) to handle the bursty requests. However, GCP-Serverless always creates much more instances than AWS-Serverless. 
For example, for VGG on GCP-Serverless, we see around 100 instances created during the first request peak, while only 50 instances are needed during the second request peak (i.e., timestamp 5-6 minutes).}
In other words, many instances are over-provisioned. This problem is moderate on AWS-Serverless. Hence, the cost of AWS-Serverless is lower, and the advantage is more obvious when the model is more complex, or the workload is higher.

\begin{figure}[t]
\begin{subfigure}[b]{0.23\textwidth}
    \centering
    \includegraphics[width=0.98\columnwidth]{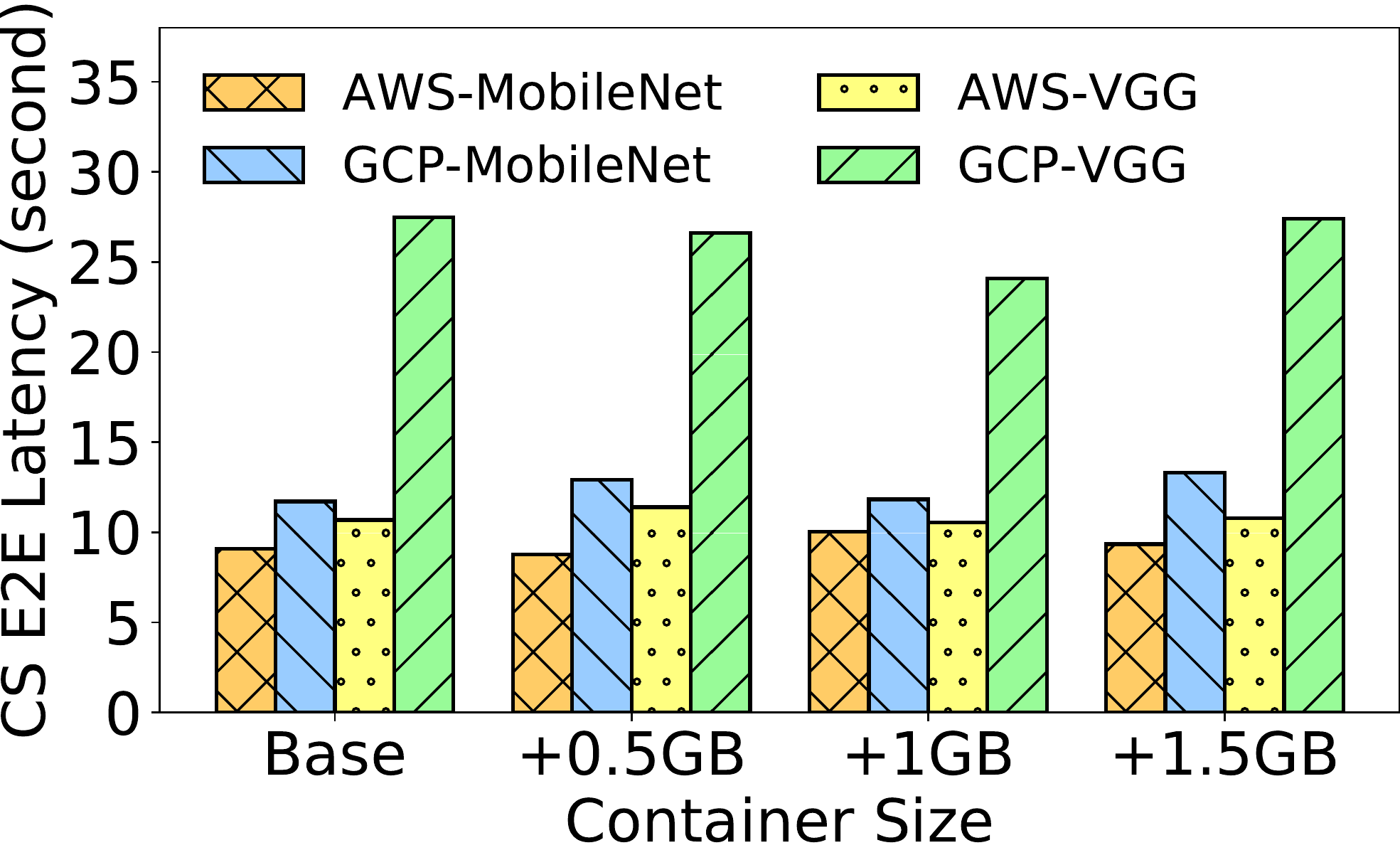}
    \vspace{-2mm}
    \caption{Vary container size}
    \label{subfig:vary-container-size}
\end{subfigure}
~
\begin{subfigure}[b]{0.23\textwidth}
    \centering
    \includegraphics[width=0.98\columnwidth]{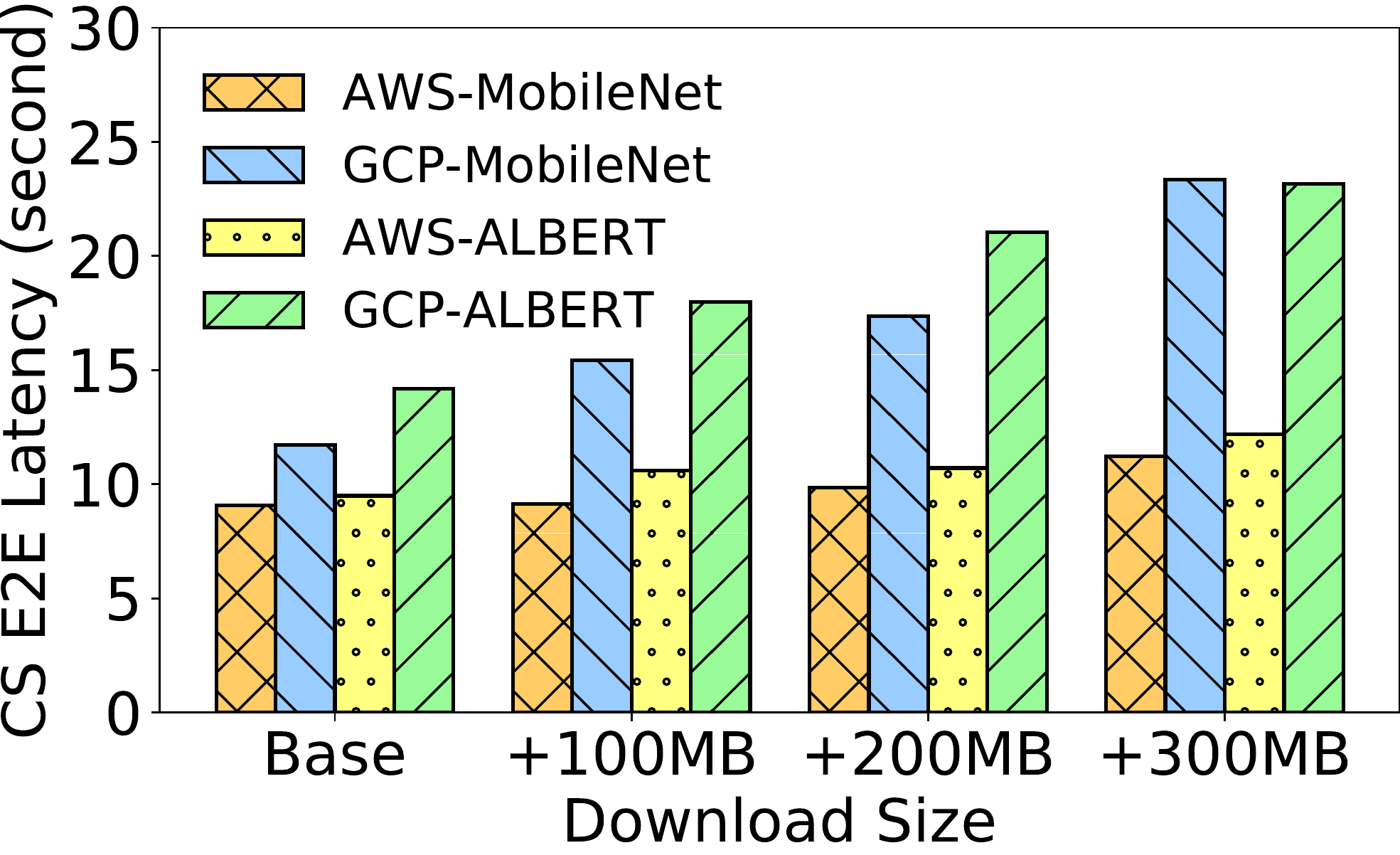}
    \vspace{-2mm}
    \caption{Vary downloaded size}
    \label{subfig:vary-download-size}
\end{subfigure}

\begin{subfigure}[b]{0.23\textwidth}
    \centering
    \includegraphics[width=0.98\columnwidth]{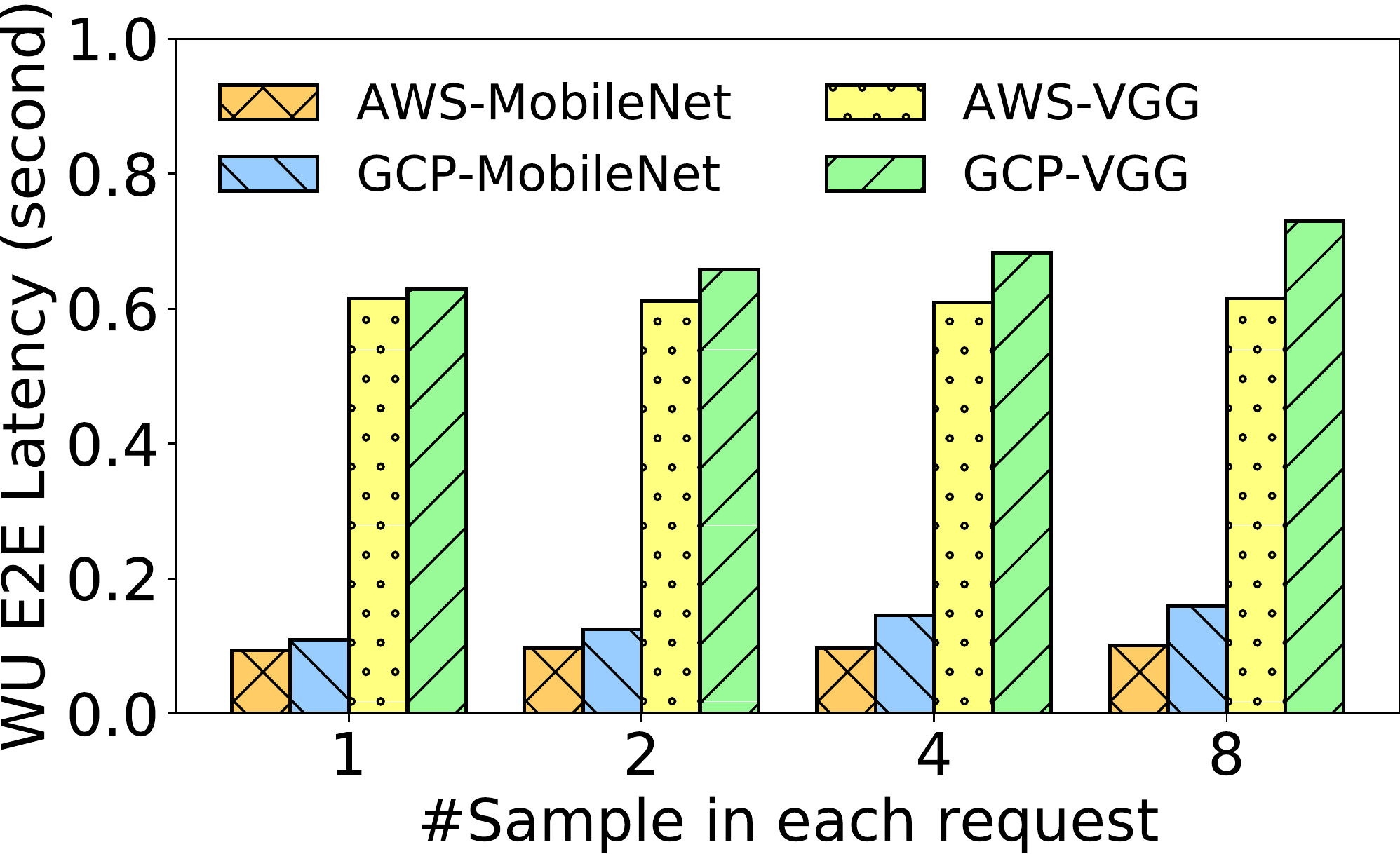}
    \vspace{-2mm}
    \caption{Vary number of samples}
    \label{subfig:vary-input-size}
\end{subfigure}
~
\begin{subfigure}[b]{0.23\textwidth}
    \centering
    \includegraphics[width=0.98\columnwidth]{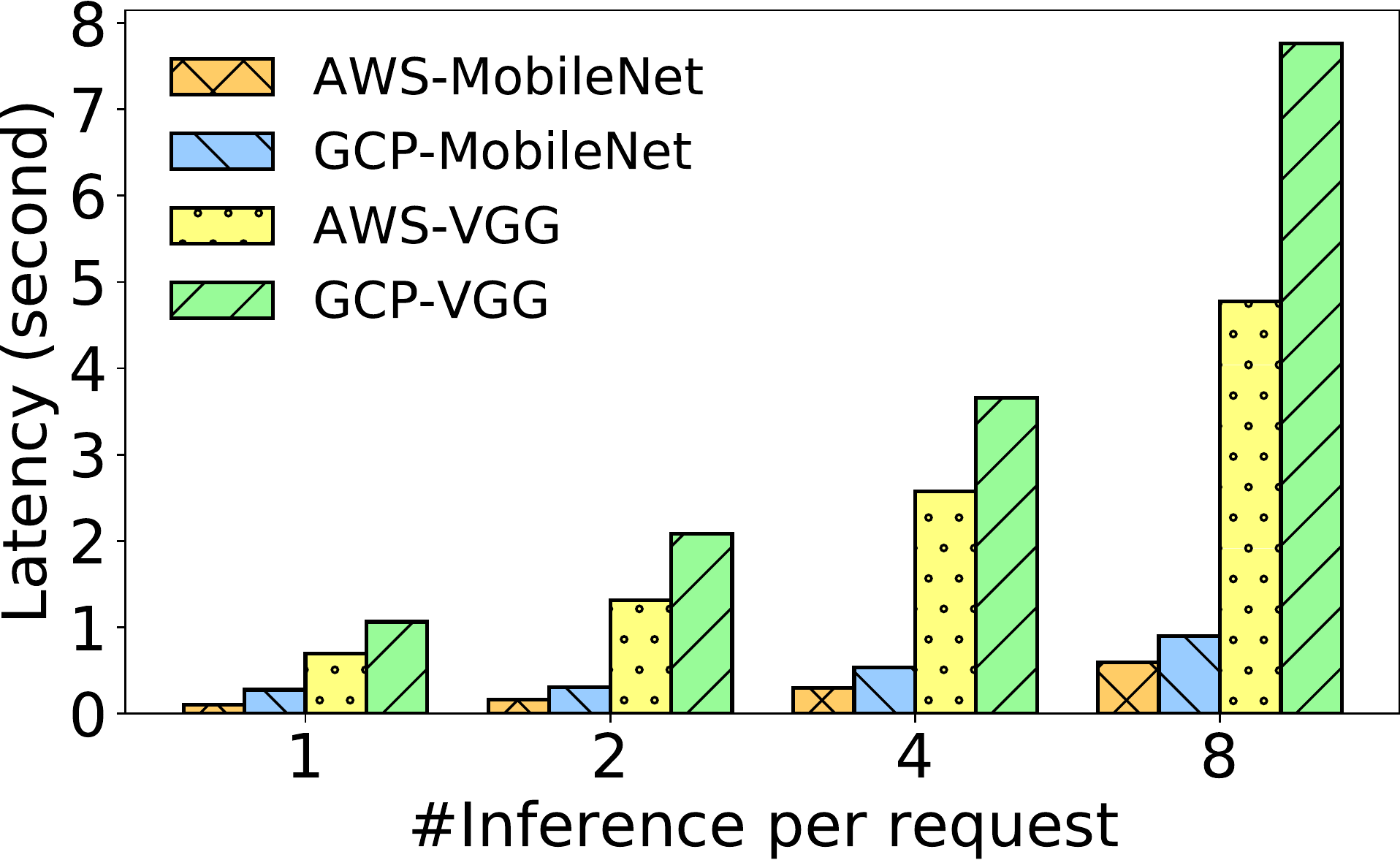}
    \vspace{-2mm}
    \caption{Vary number of inferences}
    \label{subfig:vary-inference}
\end{subfigure}
\vspace{-4mm}
\caption{In-depth analysis with workload-120.}
\vspace{-3mm}
\label{fig:in-depth-analysis}
\end{figure}

{
\noindent
\textbf{In-depth analysis.} We further conduct a set of micro-benchmark experiments with workload-120 to study the behaviors of serverless platforms. Specifically, we fix other sub-stages whenever possible and tune various inputs that may affect the performance. 

First, Figure~\ref{subfig:vary-container-size} shows the results for varying container sizes. We inject dummy files or dependencies with different sizes into the base images on both platforms. The base images are 1238MB on AWS-Serverless and 920MB on GCP-Serverless. 
We find that container size does not affect the E2E cold-start latency significantly. For instance, for MobileNet on GCP-Serverless, 
%
if we eliminate the impact of other sub-stages (i.e., import, download, load, and predict), the average remaining time of the base, base+0.5GB, base+1.0GB, and base+1.5GB containers are 1.213$s$, 1.303$s$, 1.267$s$, and 1.386$s$, respectively. This time includes the request and result transmission through the network and running the instance via the docker container. Nevertheless, the change is only 0.1$s$-0.2$s$ on average. 
%
After careful analysis, we found that only around 1\%-2\% E2E cold-start latency is much larger than the others. For example, 9 out of 738 cold-start requests consume more than 20$s$ 
for MobileNet with the base image. 
We speculate that the reason could be similar to that on OpenWhisk~\cite{openwhisk}. 
That is, the platform takes longer to run the first instance on a physical machine, i.e., it needs to pull the image from cloud storage. While for subsequent instances started on the same machine, the platform can run directly without pulling~\cite{YuLDXZLYQ020}. 

Second, we evaluate the impact of downloaded size by downloading extra dummy data besides the serving model, as shown in Figure~\ref{subfig:vary-download-size}. The E2E cold-start latency goes up as the downloaded size increases on both platforms. However, AWS-Serverless is much faster than GCP-Serverless regarding the download time, and the gap is more significant given larger dummy data. For instance, AWS-Serverless takes an extra 2.39$s$ to download 300MB of dummy data besides the ALBERT model, while GCP-Serverless takes 10.06$s$. This shows AWS-Serverless' higher downloading performance.

Third, Figure~\ref{subfig:vary-input-size} illustrates the results for varying input sizes. Specifically, we pack more samples in each client request but only predict one sample inside the function (i.e., fix the prediction sub-stage). We can see that a larger input size can slightly increase the E2E warm-up latency as more data is transmitted through the network. However, its effect on the E2E warm-up latency is minor. 

Fourth, we evaluate the impact of the predict sub-stage. We fix one sample in each request but execute the inference multiple times inside the function. Figure~\ref{subfig:vary-inference} reports the results. We observe that the overall latency grows significantly when the number of inferences increases. This is expected because the prediction sub-stage affects both cold-start and warm-up requests, and it dominates the overall latency when the predict time is large. 

\noindent
\textbf{Takeaway:} currently, AWS-Serverless' performance is better than that of GCP-Serverless. Data scientists could select AWS-Serverless as the primary platform for performant and cost-effective serving, especially with a complex model or under a high workload. Furthermore, among the factors that may affect the platforms' performance, the downloaded size and predict time are more significant.
}

\begin{figure}[t]
\centering
\begin{subfigure}[b]{0.48\textwidth}
    \hspace{2mm}
    \includegraphics[width=0.98\columnwidth]{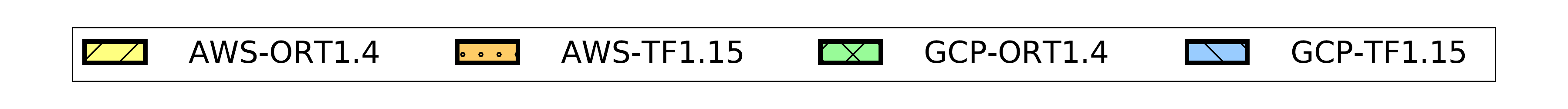}
\end{subfigure}

\begin{subfigure}[b]{0.23\textwidth}
    \centering
    \includegraphics[width=0.98\columnwidth]{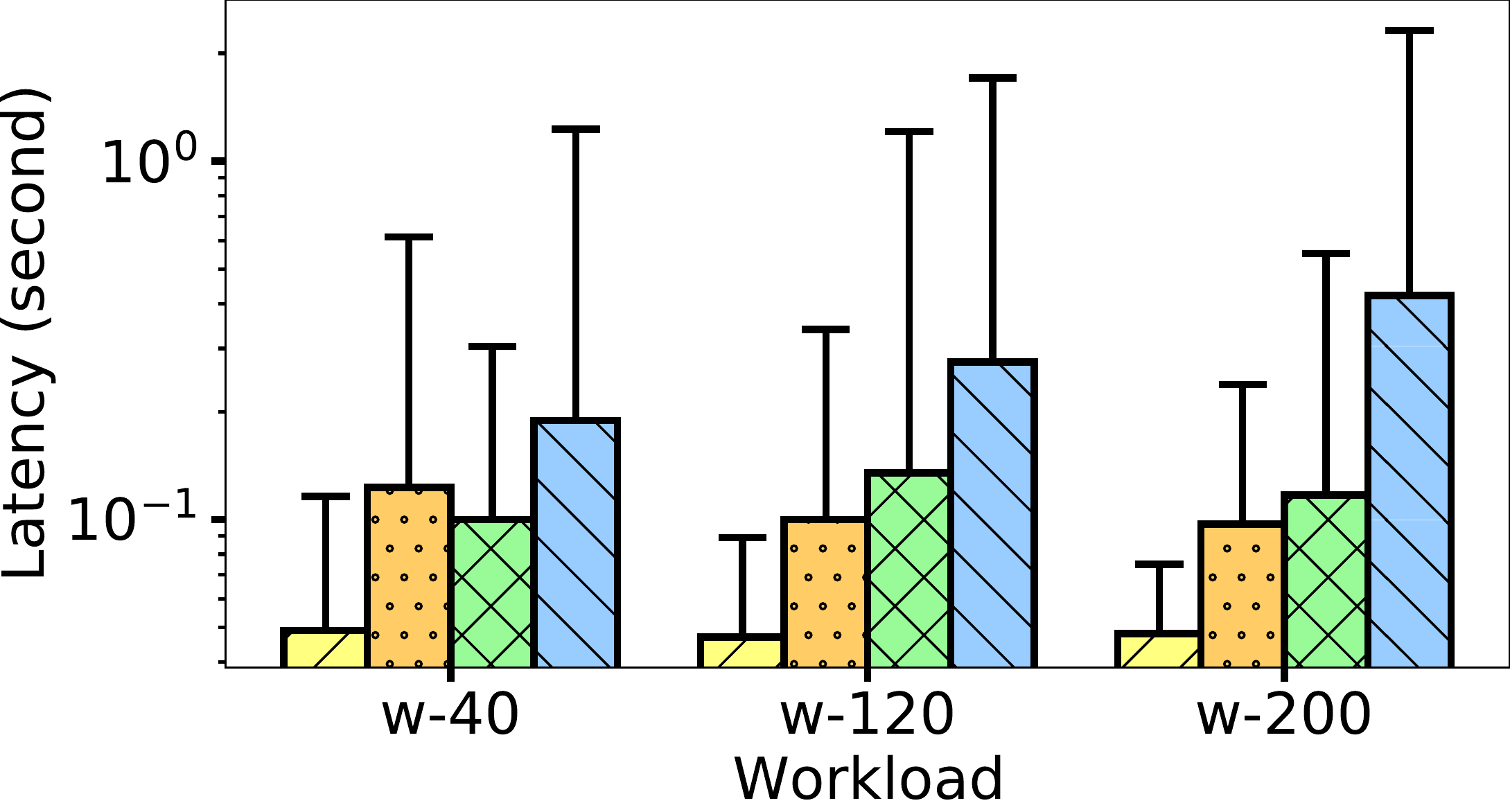}
    \vspace{-2mm}
    \caption{{MobileNet}}
    \vspace{-1mm}
    \label{subfig:runtime-mobilenet}
\end{subfigure}
~
\begin{subfigure}[b]{0.23\textwidth}
    \centering
    \includegraphics[width=0.98\columnwidth]{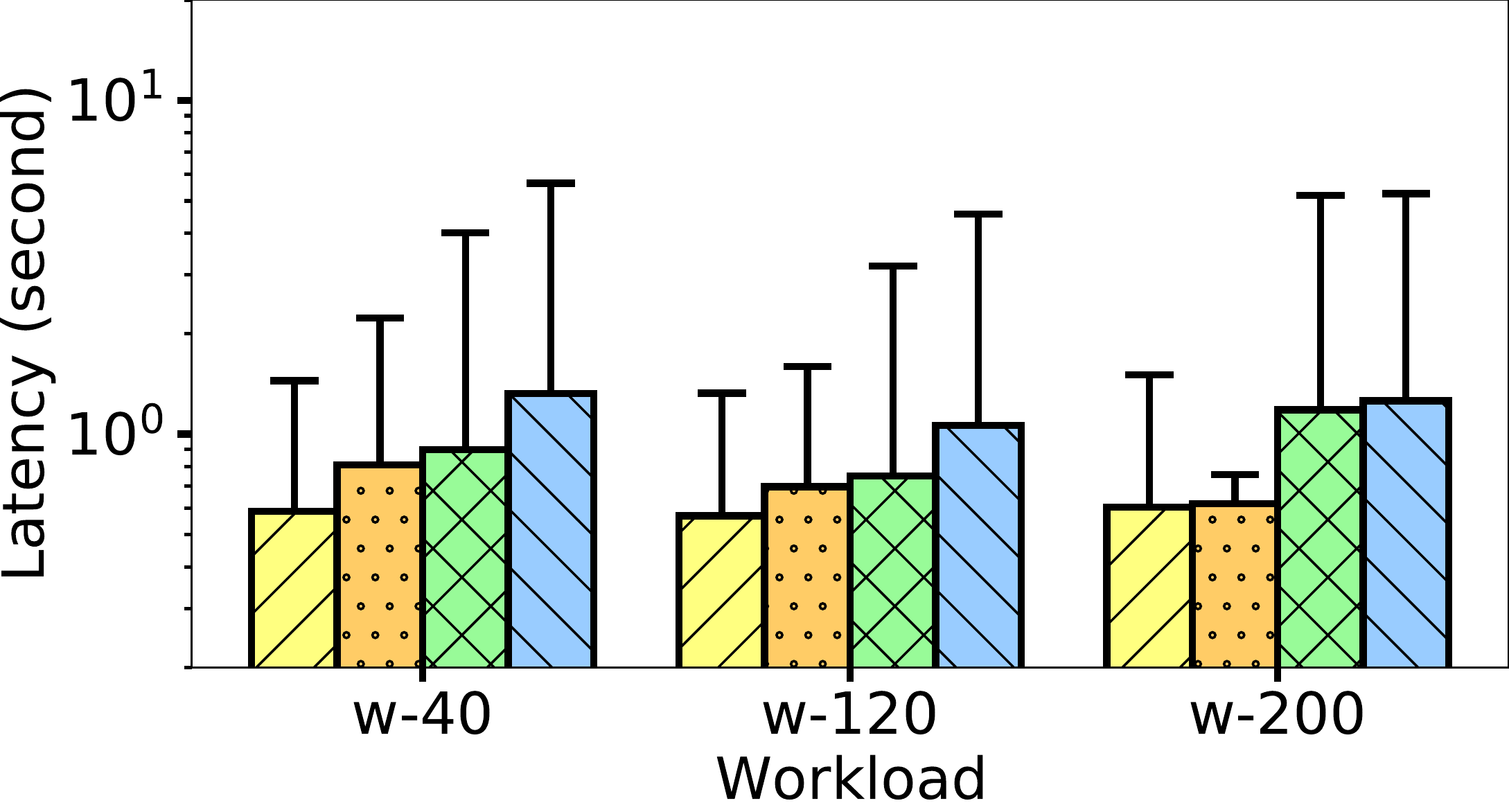}
    \vspace{-2mm}
    \caption{VGG}
    \vspace{-1mm}
    \label{subfig:runtime-vgg}
\end{subfigure}
\vspace{-2mm}
\caption{{Runtime comparison: latency w.r.t. workloads, the error bar denotes the standard deviation of latency.}}
\vspace{-3mm}
\label{fig:serverless-comparison}
\end{figure}

\begin{figure}[t]
    \centering
    \includegraphics[width=0.4\textwidth]{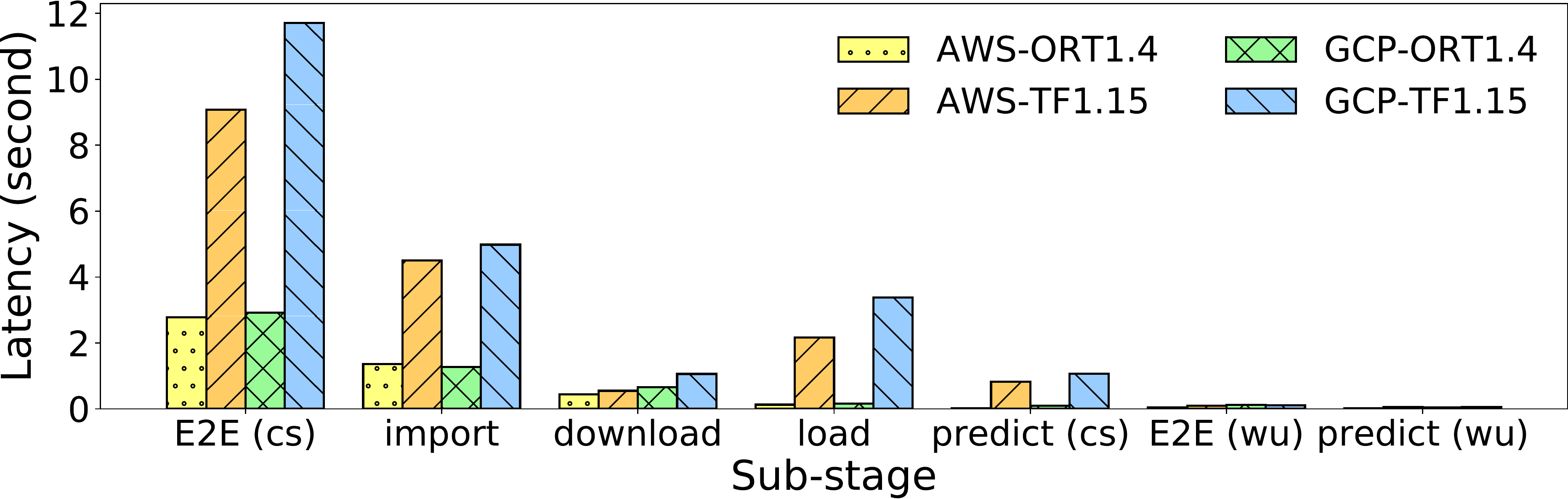}
    \vspace{-4mm}
\caption{{Breakdown comparison of different runtimes.}}
\vspace{-3mm}
\label{fig:serverless-runtime-breakdown-comparison}
\end{figure}

\ignore{
Lambda's performance is more stable than that of CF under different workloads. Even for the high workload, Lambda can spawn new instances quickly enough to handle the high number of requests. 
%
On CF, however, the longer cold-start time leads to two problems: high response latency, and over-provisioning (i.e., creating more instances than needed).
Specifically, for TF1.15 under workload-200 (see Figure~\ref{subfig:workload-200}), we see more than 1200 instances created during the first request peak (i.e., timestamp 0 to 50), while only 500 instances are needed during the second request peak (i.e., timestamp 150 to 200). In other words, about 700 instances are over-provisioned due to the high request rate. 
As a result, the cost becomes higher, as shown in Table~\ref{table:serverless-costs-comparison}. 

\textit{Still be better to add a subsection introduce the main differences on Lambda and CF. Emphasize three points: (1) different cold start time (together with the trend that workload-40 average latency larger than workload-200 for ALBERT and VGG); (2) different number of instances created with Figure~\ref{fig:serverless-vs-mlservice-instance-num}; (3) different success ratios (may due to cold start time); (4) different costs for the experiments.}
}

\subsection{Serving Runtime}
\label{subsec:runtime}


Then, we evaluate the impact of serving runtime. Specifically, we compare OnnxRuntime1.4 (ORT1.4) with TensorFlow1.15 (TF1.15). 
%
%
Figure~\ref{fig:serverless-comparison} and Table~\ref{table:runtime-cost-comparison} report the average overall latency and cost for MobileNet and VGG, respectively. 
There are two main observations. 

First, model serving with ORT1.4 is more efficient than with TF1.15 on both platforms.
It has up to 2.51$\times$ lower in latency and 4.55$\times$ lower in cost for AWS-Serverless, and 3.61$\times$ and {1.97$\times$} for GCP-Serverless, respectively. 
%
The reason is two-fold. One is that the E2E cold-start latency is significantly reduced. Figure~\ref{fig:serverless-runtime-breakdown-comparison} details the breakdown comparison between the two runtimes for MobileNet with workload-120. Specifically, the E2E cold-start latency drops from 9.08$s$ to 2.775$s$ on AWS-Serverless, and from 11.71$s$ to 2.917$s$ on GCP-Serverless.
Note that the container size for ORT1.4 is 391MB for MobileNet on AWS-Serverless, compared to 1238MB for TF1.15. Though the container size has little effect on the cold-start latency (as discussed in Section~\ref{subsec:serverless-comparison}), we observe that using ORT1.4 greatly reduces the import time and load time. 
%
The other reason is that ORT1.4 is optimized for improved model inference, whereas TF1.15 is the standard runtime with a relatively longer execution time per request. 
For example, if analyzing the predict time of warm-up requests for MobileNet on GCP-Serverless, the average predict time is 0.061$s$ for TF1.15, while that for ORT1.4 is 0.043$s$. 

Second, the improvement for using ORT1.4 on MobileNet is more significant than on VGG. 
For instance, for MobileNet on GCP-Serverless, ORT1.4 is up to 3.61$\times$ faster than TF1.15 and 1.97$\times$ lower in cost, while for VGG, the improvements are 1.47$\times$ and 1.32$\times$, respectively.
The rationale is that the cold-start time dominates the average latency and cost for the MobileNet model as the actual execution time per request is very short. While for the VGG model, the actual execution time is longer, which has a relatively large impact on the performance and cost, leading to a moderate improvement. 

\noindent
\textbf{Takeaway:} {data scientists should consider a smaller and faster serving runtime (e.g., ORT) on serverless platforms as the first choice as long as that runtime supports the desired model.} This can greatly improve the performance and save cost.

\begin{table}[t]
\centering
\scriptsize
\caption[]{Costs for serverless serving with ORT1.4}
\vspace{-4mm}
\begin{tabular}{| c | c | c | c | c |}
\hline
\multicolumn{2}{|c|}{{Serving Services}} & workload-40 & workload-120 & worload-200 \\
\hline
\hline
\multirow{2}{*}{{{AWS-Serverless}}} & {MobileNet} & {\$0.011} & {\$0.037} & {\$0.062} \\
& {VGG} & {\$0.322} & {\$0.931} & {\$1.644} \\
\hline
\multirow{2}{*}{{{GCP-Serverless}}} & {MobileNet} & {\$0.047} & {\$0.160} & {\$0.272} \\
& {VGG} & {\$0.383} & {\$1.108} & {\$2.455} \\
\hline
\end{tabular}
\label{table:runtime-cost-comparison}
\vspace{-3mm}
\end{table}

\subsection{Memory Size}
\label{subsec:memory}

\begin{figure}[t]
\centering
\begin{subfigure}[b]{0.48\textwidth}
    \hspace{-1mm}
    \includegraphics[width=0.98\columnwidth]{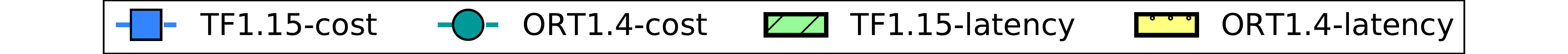}
\end{subfigure}

\begin{subfigure}[b]{0.23\textwidth}
    \centering
    \includegraphics[width=0.98\columnwidth]{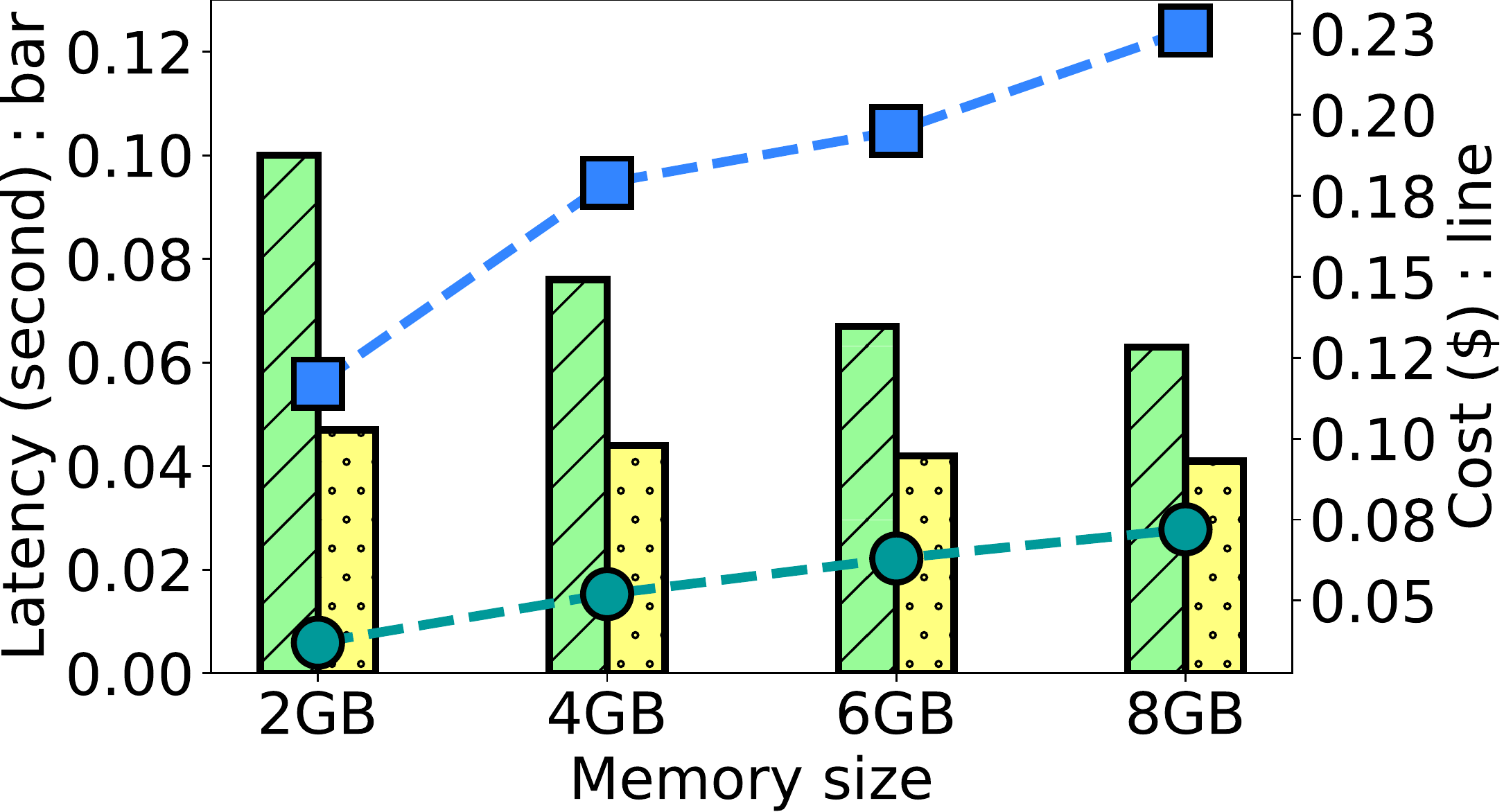}
    \vspace{-2mm}
    \caption{MobileNet with w-120}
    \vspace{-1mm}
    \label{subfig:memory-size-mbnet}
\end{subfigure}
~
\begin{subfigure}[b]{0.23\textwidth}
    \centering
    \includegraphics[width=0.98\columnwidth]{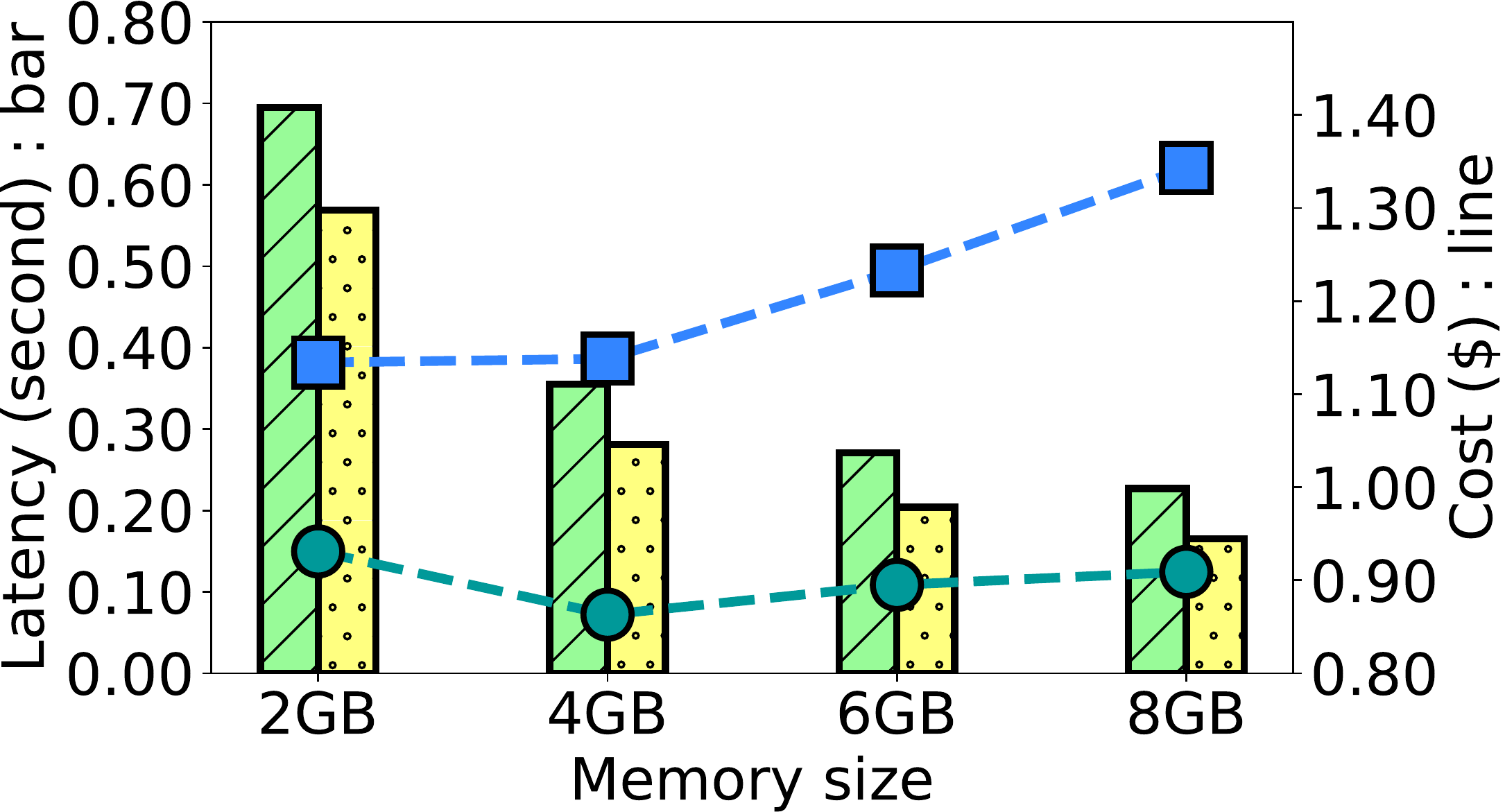}
    \vspace{-2mm}
    \caption{VGG with w-120}
    \vspace{-1mm}
    \label{subfig:memory-size-vgg}
\end{subfigure}
\vspace{-2mm}
\caption{Vary memory size on AWS-Serverless.}
\vspace{-3mm}
\label{fig:memory-size}
\end{figure}

Next, we investigate the impact of several function-specific parameters on the performance of serverless model serving on AWS-Serverless. The first is memory size. Figure~\ref{fig:memory-size} shows the average latency and cost of MobileNet and VGG using TF1.15 and ORT1.4 with workload-120. For both models, the latency decreases as memory size increases because each request can be processed faster. 
Besides, the latency decrease of the VGG model is sharper than that of MobileNet as the memory size goes up. 
{
This is because the two runtimes already execute MobileNet very fast given 2GB, so that increasing memory cannot improve the performance as significantly as the VGG model.
%
%
In addition, the predict time is less dominant in the end-to-end latency, e.g., for MobileNet served with ORT1.4, the average predict time is about 0.012$s$ given 2GB, whereas the overall latency is about 0.047$s$, rendering its improvement (e.g., 0.009$s$ given 4GB) insignificant to the overall performance.} 

Another interesting phenomenon is that a larger memory does not always increase the cost. For example, for the VGG model in Figure~\ref{subfig:memory-size-vgg}, we observe that the 4GB memory can even slightly reduce the cost. There are two reasons. One is that a larger memory decreases the execution time per request due to the improved processing capacity (as discussed above). The other is that it also reduces the number of cold-started instances. 
For instance, by increasing 2GB to 4GB for VGG with ORT1.4, the number of cold-started instances is decreased from 408 to 37 in our experiment. 
%
However, when we keep increasing the memory size to 6GB or 8GB, the cost does not reduce further as the two improvements cannot compensate for the additional cost charged for larger memory.

\noindent
\textbf{Takeaway:} if latency is the primary consideration, data scientists should choose a large memory. Besides, data scientists could run several warm-up requests beforehand and check the proportion of predict time to end-to-end latency. If the proportion is minor, they can use a small memory; otherwise, they could select a relatively large memory. 
A more sophisticated way is to use a memory tuning tool (e.g., \cite{lambda-power-tuning}) to find an optimized size; nevertheless, it requires data scientists to have a better understanding of the workload.


\subsection{Provisioned Concurrency} \label{subsec:concurrency}

\begin{figure}[t]
\centering
\begin{subfigure}[b]{0.48\textwidth}
    \hspace{-1mm}
    \includegraphics[width=0.98\columnwidth]{figures/revision/revision-legend-investigation.pdf}
\end{subfigure}

\begin{subfigure}[b]{0.23\textwidth}
    \centering
    \includegraphics[width=0.98\columnwidth]{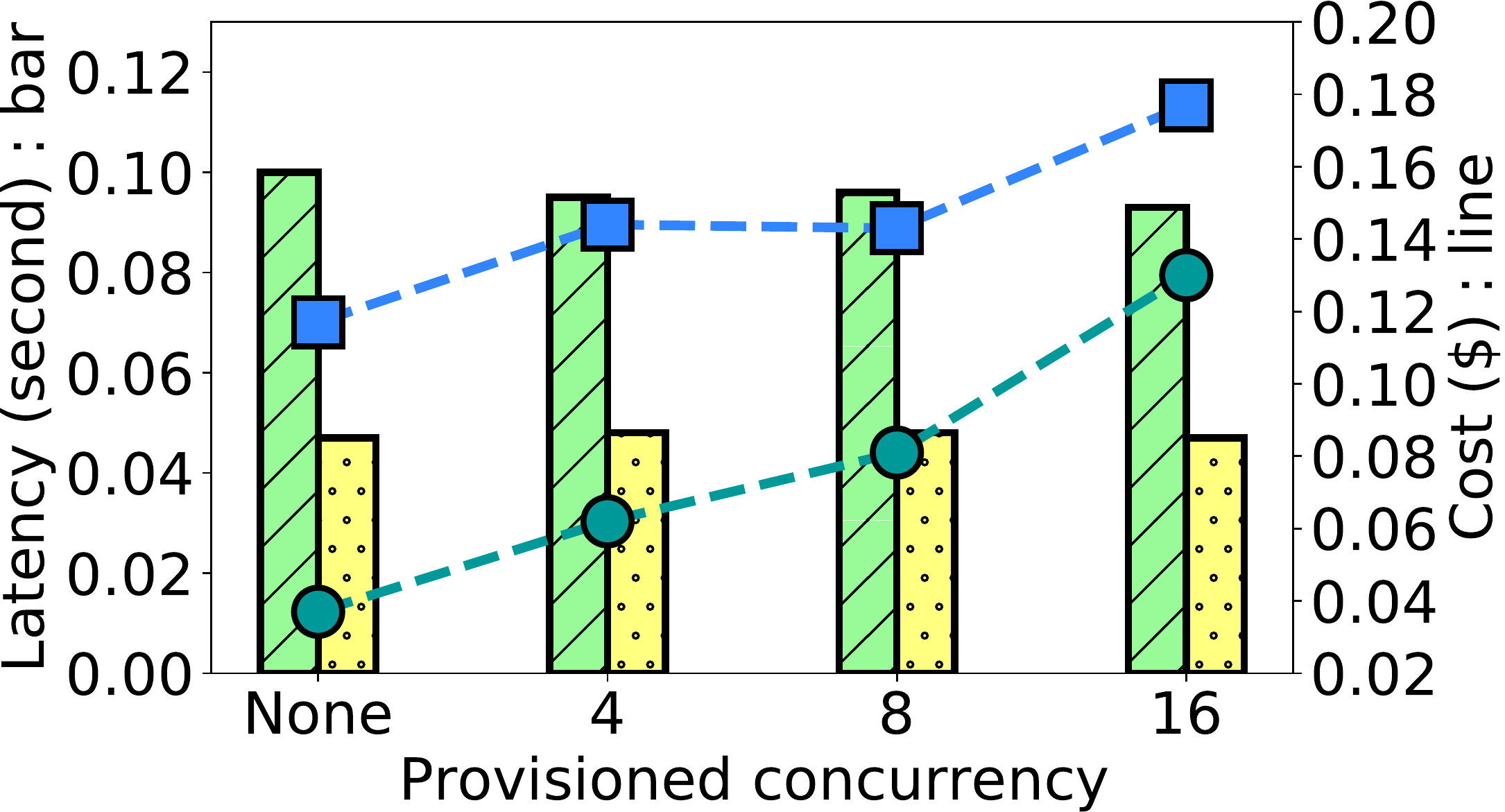}
    \vspace{-2mm}
    \caption{MobileNet with w-120}
    \vspace{-1mm}
    \label{subfig:concurrency-mbnet}
\end{subfigure}
~
\begin{subfigure}[b]{0.23\textwidth}
    \centering
    \includegraphics[width=0.98\columnwidth]{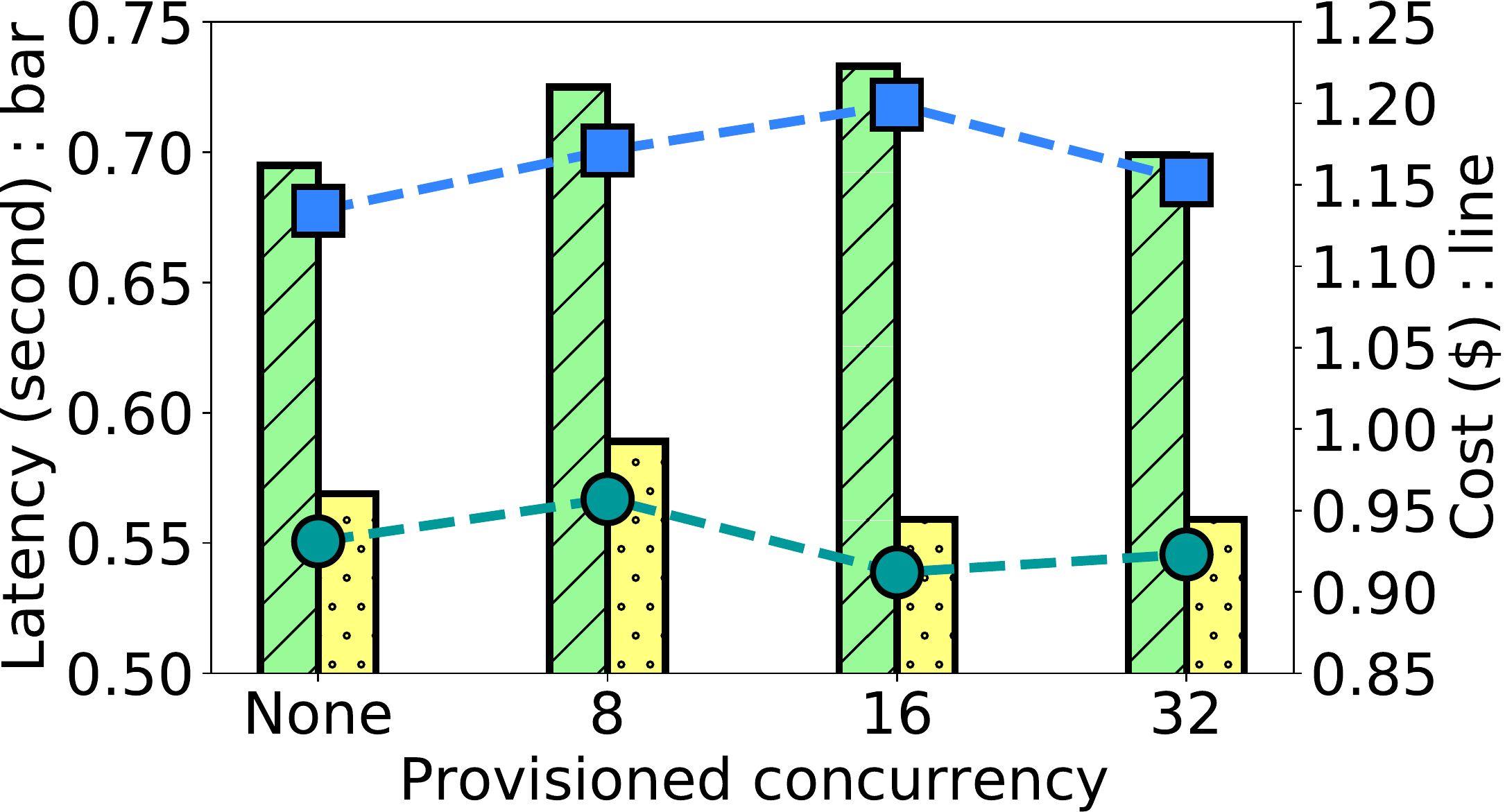}
    \vspace{-2mm}
    \caption{VGG with w-120}
    \vspace{-1mm}
    \label{subfig:concurrecy-vgg}
\end{subfigure}
\vspace{-2mm}
\caption{Vary provisioned concurrency on AWS-Serverless}
\vspace{-3mm}
\label{fig:concurrency}
\end{figure}

We now study the impact of provisioned concurrency on AWS-Serverless, i.e., a number of instances are always keeping warm during the workload. Basically, this is a hybrid approach that combines serverless and server-based model serving. Figure~\ref{fig:concurrency} reports the performance and cost for MobileNet and VGG with workload-120. 
We observe that provisioned concurrency does not necessarily reduce average latency, and for the VGG model, the latency is even higher in some cases. 
This seems counter-intuitive as a portion of the requests can be forwarded to the provisioned instances and processed quickly, which should reduce the average latency. However, we found that the number of cold-started instances sometimes increases given provisioned concurrency. 
For example, for VGG serving with TF1.15, the number of cold-started instances are 614, 640, and 478, for provisioned concurrency 8, 16, and 32, respectively, compared to 409 for no provisioned concurrency.
Since AWS-Serverless' scaling strategy is a black-box to users, we speculate that it adopts a more aggressive scaling policy when using provisioned concurrency, i.e., once it finds that the provisioned instances are insufficient, it scales more instances to handle the rest of requests than that without provisioned concurrency. 

\noindent
\textbf{Takeaway:} the current provisioned concurrency method may not be a good choice for serverless model serving. A related suggestion to cloud providers is to publicize their autoscaling policy so that users can better utilize it to improve performance and cost.

\subsection{Batch Size}
\label{subsec:batch-size}

\begin{figure}[t]
\centering
\begin{subfigure}[b]{0.48\textwidth}
    \hspace{-1mm}
    \includegraphics[width=0.98\columnwidth]{figures/revision/revision-legend-investigation.pdf}
\end{subfigure}

\begin{subfigure}[b]{0.23\textwidth}
    \centering
    \includegraphics[width=0.98\columnwidth]{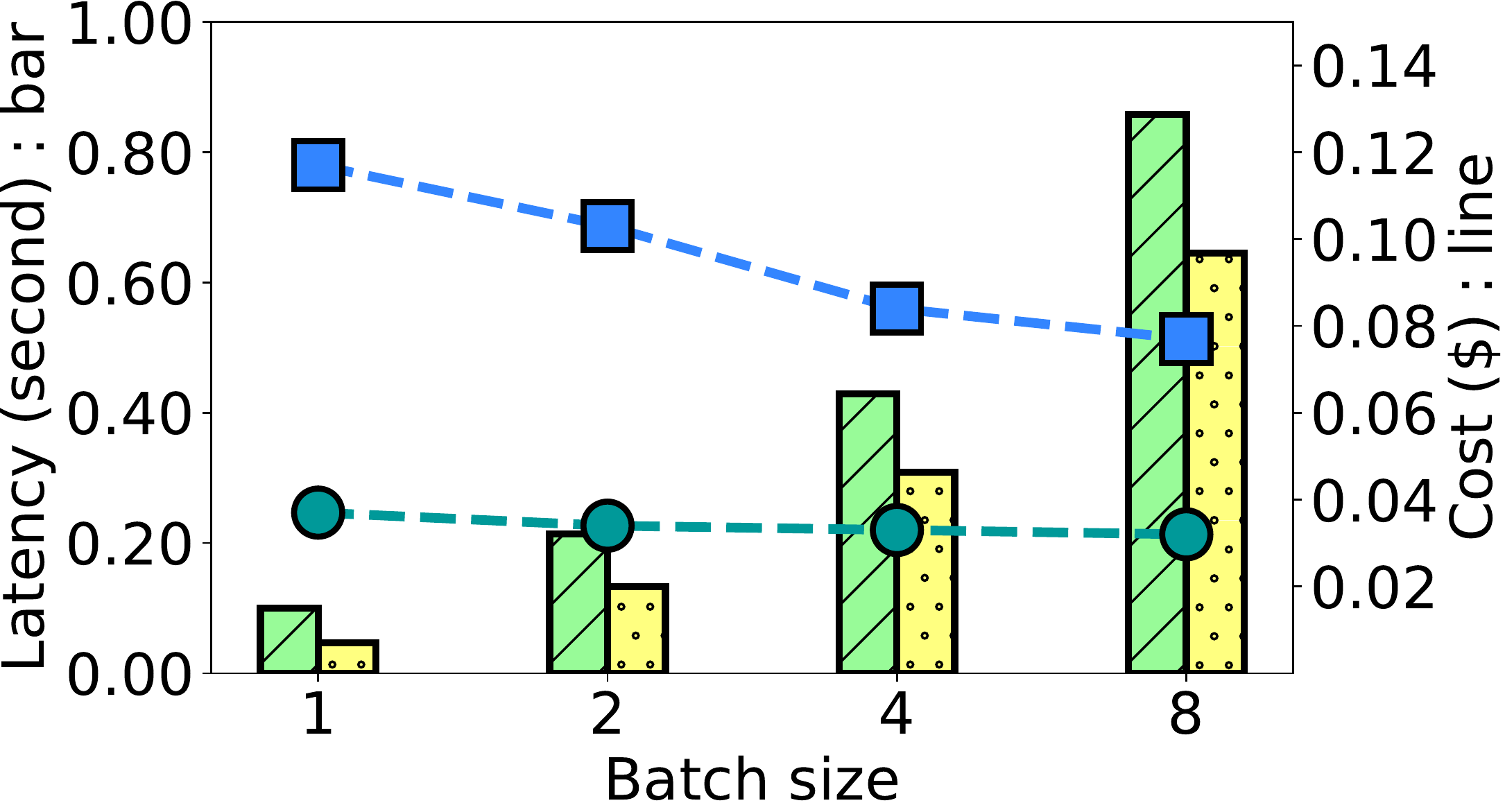}
    \vspace{-2mm}
    \caption{MobileNet with w-120}
    \vspace{-1mm}
    \label{subfig:batch-size-mbnet}
\end{subfigure}
~
\begin{subfigure}[b]{0.23\textwidth}
    \centering
    \includegraphics[width=0.98\columnwidth]{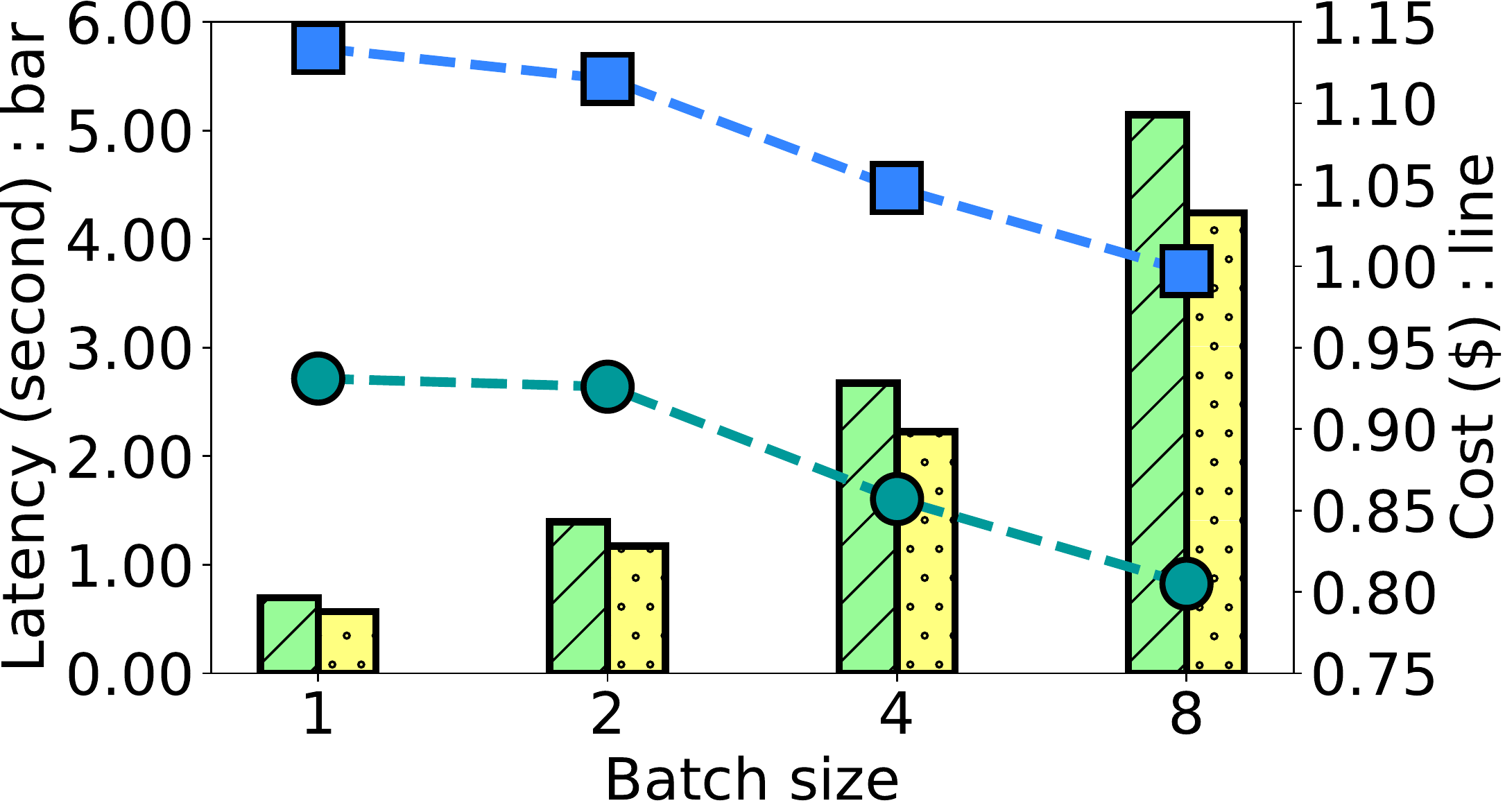}
    \vspace{-2mm}
    \caption{VGG with w-120}
    \vspace{-1mm}
    \label{subfig:batch-size-vgg}
\end{subfigure}
\vspace{-2mm}
\caption{Vary batch size on AWS-Serverless.}
\vspace{-3mm}
\label{fig:batch-size}
\end{figure}

Finally, we investigate the impact of batch size on the performance and cost of AWS-Serverless. Given a batch size, each client sends an invocation to the serverless function only when the number of requests matches the batch size or reaches the end of the workload. Figure~\ref{fig:batch-size} presents the results for MobileNet and VGG with workload-120. There are two main observations. First, the average latency is approximately doubled when the batch size doubles. This is because most requests are delayed, and the execution time on serverless (i.e., response to the batched requests) grows, which increases the latency. Second, in most cases, batching could reduce the cost because: (1) the number of invocations is reduced; and (2) the request rate is reduced, resulting in a fewer number of cold-started instances. However, for MobileNet with ORT1.4, the cost reduction is insignificant since the model is simple and ORT1.4 is already able to handle requests effectively with very few instances. 

\noindent
\textbf{Takeaway:} if cost is the primary concern, data scientists could batch requests; 
otherwise, batching is not suggested as the response latency will increase significantly.
A better way is to apply an adaptive batching strategy given the cost and latency constraints~\cite{BATCH-serve}, but it requires a careful design to predict the request rate.

\section{Challenges and Opportunities}\label{sec:discussion}

Though existing serverless platforms have provided full-fledged functionalities that make serverless a viable model serving option for data science applications, there are still several research challenges and opportunities for serverless model serving.

The first challenge is how to address the over-provisioning problem~\cite{Ustiugov0KBG21, FuerstS21}, which can greatly increase average latency and cost (see Section~\ref{subsec:serverless-comparison}). 
%
This problem can be exacerbated when serving increasingly larger models or under higher workloads.
Therefore, designing new scaling policies for serverless model serving is an important and promising research direction. Efficient policies may monitor the requests' execution time, predict the subsequent request rate, and allocate the needed instances precisely. 
%

The second challenge is the protection of data security. With strict privacy regulations~\cite{GDPR2016}, data scientists now need to consider user's data security. 
%
%
A possible solution is based on trusted hardware~\cite{McKeenABRSSS13, ZhengDBPGS17}, to ensure secure computation. Unfortunately, most existing serverless platforms do not support this functionality. Besides, the data protection should be supported in a scalable manner to accommodate serverless model serving.
%
%
Thus, more research work could be conducted to make serverless secure and efficient.

The third challenge is the complexity of the design space for serverless model serving. As investigated in Section~\ref{sec:investigation}, many factors affect both the performance and cost. 
The consequence is that the data scientists have more decisions to make, which affects their productivity. A potential direction is to build a navigation tool that automatically searches the design space for serverless deployment, and finds the best configuration under pre-defined constraints, such that data scientists can
adopt serverless model serving effectively.


\ignore{
The third challenge is that deep learning models are becoming larger and more complex, which may lead to significantly high response latency. 
Though parallelism is a possible strategy to reduce the latency, the limited resources on serverless instances make it not efficient enough. 
%
Besides, some models may even not fit into the serverless instances due to the limited memory size~\cite{mark}. A possible direction is to use multiple instances to execute one model prediction in a distributed manner. 
However, more research work should be conducted to investigate how to determine a reasonable model slicing strategy and parallelize the executions effectively, for reducing latency and cost.
}
\section{Conclusions}\label{sec:conclusion}

In this paper, we conduct a comprehensive comparison of serverless against other model serving systems from AWS and GCP.
%
The evaluation results demonstrate that serverless is a viable and promising option for model serving. 
%
%
We further investigate the design space of serverless model serving regarding various choices and present recommendations for data scientists to better utilize serverless. Finally, we discuss challenges and opportunities in building a more practical serverless model serving system. 

\section*{ACKNOWLEDGEMENTS}\label{sec:acknowledgements}

We thank 
Pan Zhou for his early contribution to this work. This research is supported by Singapore Ministry of Education Academic Research Fund Tier 3 under MOEs official grant number MOE2017-T3-1-007. Tien Tuan Anh Dinh's work is supported by Singapore University of Technology and Design's startup grant SRG-ISTD-2019-144. Meihui Zhang's work is supported by National Natural Science Foundation of China (62072033). 

\balance
\bibliographystyle{ACM-Reference-Format}
\bibliography{ref}

\end{document}